\documentclass[fleqn,usenatbib]{mnras}

\usepackage{newtxtext,newtxmath}

\usepackage[T1]{fontenc}
\usepackage{ae,aecompl}
\usepackage{times}

\usepackage{graphicx}	
\usepackage{amsmath}	
\usepackage{amssymb}	

\newcommand{\bs}{\boldsymbol}

\newcommand{\percent}{\,\mathrm{per cent}}

\newcommand{\levicivita}[3]{\epsilon_{#1#2#3}}

\title[
The pattern speed from transverse velocities 
]{
The pattern speed of the Milky Way bar from transverse velocities
}

\author[J. L. Sanders et al.]{
Jason L. Sanders,$^{1}$\thanks{E-mail: jls@cam.ac.uk (JLS), nwe@ast.cam.ac.uk (NWE)}
Leigh Smith,$^{1,2}$
N. Wyn Evans$^{1}$
\\
$^{1}$Institute of Astronomy, University of Cambridge, Madingley Rd, Cambridge, CB3 0HA, UK\\
$^{2}$School of Physics, Astronomy and Mathematics, University of Hertfordshire, College Lane, Hatfield AL10 9AB, UK\\
}

\date{Accepted XXX. Received YYY; in original form ZZZ}

\pubyear{2019}

\begin{document}
\label{firstpage}
\pagerange{\pageref{firstpage}--\pageref{lastpage}}
\maketitle

\begin{abstract}
We use the continuity equation to derive a method for measuring the pattern speed of the Milky Way's bar/bulge from proper motion data. The method has minimal assumptions but requires complete coverage of the non-axisymmetric component in two of the three Galactic coordinates. We apply our method to the proper motion data from a combination of Gaia DR2 and VISTA Variables in the Via Lactea (VVV) to measure the pattern speed of the bar as $\Omega_\mathrm{p}=(41\pm 3)\,\mathrm{km\,s^{-1}\,kpc^{-1}}$ (where the error is statistical). 
This puts the corotation radius at 
$(5.7\pm0.4)\,\mathrm{kpc}$, 
under the assumptions of the standard peculiar motion of the Sun and the absence of non-axisymmetric streaming in the Solar neighbourhood.
The obtained result uses only data on the near-side of the bar which produces consistent measurements of the distance and velocity of the centre of the Galaxy. Addition of the data on the far-side of the bar pulls the pattern speed down to $\Omega_\mathrm{p}=(31\pm 1)\,\mathrm{km\,s^{-1}\,kpc^{-1}}$ but requires a lower transverse velocity for the Galactic centre than observed. This suggests systematics of $5-10\,\mathrm{km\,s^{-1}kpc^{-1}}$ dominate the uncertainty.
We demonstrate using a dynamically-formed bar/bulge simulation that even with the limited field of view of the VVV survey our method robustly recovers the pattern speed.
\end{abstract}




\section{Introduction}

The pattern speed of the central bar of the Milky Way is a fundamental parameter for characterising our Galaxy. As the orbits that support a bar do not exist much beyond corotation, the length of a bar is set by its pattern speed~\citep{Ag98}. We naturally wish to understand how our Galaxy is structured and how it compares to other barred spirals~\citep{BlandHawthornGerhard2016}. The pattern speed is an essential parameter that allows us to make such direct comparisons~\citep{Aguerri2015,Guo2019}. Interpretation of other Galactic observations require a robust understanding of the bar and its resonances -- for example, observations of the solar neighbourhood velocity substructure~\citep{Ka91,Dehnen1999,Mo17} or interpretation of the high velocity peaks seen in radial velocity surveys towards the Galactic Centre~\citep{Mo15,AU15} or analysis of the bimodal distribution of red clump magnitudes~\citep{Na10,McW10}. The pattern speed of the bar is related to our understanding of when and how it formed, and how it has subsequently interacted with other components in the Galaxy. For instance, bars may be slowed significantly via dynamical friction through interaction with a dark matter halo~\citep{De00}, whilst buckling instabilities can transform rapidly rotating bars into more sedately rotating peanut-shaped bars~\citep{Ra91}.

The pattern speed of the bar has proved to be an awkward parameter to pin down. In part, this is because it is not trivially related to the velocity of its constituent stars, but instead describes the rate of figure rotation of the density and potential of the bar. Stars or gas within the bar possess net streaming in the bar frame, as the orbits that support a bar are preferentially prograde rotating \citep[e.g.,][]{BinneyTremaine2008}. \cite{Gerhard2011} summarises the methods used for the measurement of the pattern speed in the Milky Way. Indirect methods include use of hydrodynamical simulations to reproduce features in the Galactic longitude versus CO and HI terminal velocity ($\ell,v$) maps~\citep{Fu99,Bi03}. This relies on the fact that gas traces the closed orbital structure and so is an excellent probes of the gravitational potential of the bar. Another indirect method identifies the Hercules stellar streams in the local stellar velocity distribution with a family of resonant orbits 
in a barred potential.
Evidence for the interpretation of the Hercules stream as a resonance has mounted in recent years~\citep{My18,Hu18}, though its causation remains unclear.
There is a dichotomy between models producing the Hercules stream as the outer Lindblad resonance of a short-fast bar \citep{Dehnen1999,An14} and those with a long-slow bar that produce the Hercules stream as a corotation resonance \citep{Pe17} or a 4:1 outer Lindblad resonance \citep{Hunt2018}.
However, the most direct method was introduced by \citet{TremaineWeinberg} and is derived from the continuity equation with minimal assumptions. It has now been successfully applied to many external galaxies~\citep{Me95,Ge99,De02,Aguerri2015}. \cite{Debattista2002} derived a version of the \citet{TremaineWeinberg} method applicable to line of sight velocity datasets in the Galaxy and demonstrated its use on a sample of $\sim 700$ OH/IR stars in the bar.

\cite{BlandHawthornGerhard2016} report a number of measurements of the pattern speed. Recent hydrodynamical modelling has obtained
$\Omega_{\rm p} = 42$ $\mathrm{km\,s}^{-1}$ kpc$^{-1}$ \citep{We99}, 30-40 $\,\mathrm{km\,s}^{-1}$ kpc$^{-1}$ \citep{Co08} and 40 $\mathrm{km\,s}^{-1}$ kpc$^{-1}$ \citep{So15}. Using the Hercules stream tends to give higher values; \citet{Dehnen2000} using {\it Hipparcos} data originally found
$\Omega_{\rm p} = (51\pm 4)$ $\mathrm{km\,s}^{-1}$ kpc$^{-1}$, while more recent studies have obtained $(51.5 \pm 1.5)$ \citep{Mi07} and
$(53 \pm 0.5)$ km\,s$^{-1}$ kpc$^{-1}$ \citep{An14}. The only previous attempt to use the \citet{TremaineWeinberg} method for the Galactic bar gave one of the highest values of the pattern speed of all, namely $\Omega_{\rm p} = (59 \pm 15)\,\mathrm{km\,s}^{-1}$ kpc$^{-1}$ \citep{Debattista2002}.

Additionally, a lowish pattern speed is supported by the fully dynamical models of the stellar populations of the bar. For example, \citet{Portail2015} used Made-to-Measure methods to reproduce the three-dimensional density of red clump giants, as well as the BRAVA line of sight velocity data~\citep[e.g.,][]{Ku12} in selected fields. This concluded that a still lower pattern speed of $\Omega_{\rm p} = 25-30\,\mathrm{km\,s}^{-1}$ kpc$^{-1}$ is favoured. Subsequent hydrodynamical simulations using the same gravitational force field seem to confirm the low pattern speed as providing the best fit to date to the observed ($\ell,v$) data~\citep{Li16}.  
Since then, by incorporating additional kinematic data from the OGLE~\citep{Rattenbury2007} and ARGOS surveys~\citep{Ne16} into the Made-to-Measure modelling, \cite{Portail2017} measured a somewhat higher pattern speed of $\Omega_{\rm p} = (39\pm 3.5)\,\mathrm{km\,s}^{-1}$kpc$^{-1}$, consistent with the recommended combined estimate of \cite{BlandHawthornGerhard2016} of $\Omega_\mathrm{p} = (43 \pm 9)$ $\mathrm{km\,s}^{-1}$ kpc$^{-1}$ using  pre-2016 literature estimates.

It is apparent that, despite a lot of effort, the pattern speed of the bar has resisted an easy concensus. Given the advent of new proper motion catalogues from the Gaia satellite~\citep{Gaia2016,Gaia2018} and elsewhere~\citep{Sm18}, now
seems a propitious moment to generalise the \citet{TremaineWeinberg} method to transverse motions and apply it to the Galactic bar anew. 

In a companion paper (Sanders, Smith, Evans \& Lucas 2019, hereafter Paper I), we extracted the velocity field of the bar/bulge from the proper motions of $\sim45$ million stars across the Vista Variables in the Via Lactea \citep[VVV,][]{Minniti2010,Saito2012} survey. We used proper motions from a combination of Gaia DR2 and the VIRAC catalogue. The VIRAC catalogue used the multi-epoch data from VVV to compute relative proper motions which were fixed to an absolute frame using Gaia DR2. In this paper, we use the results of Paper I to measure the pattern speed of the bar/bulge. Our data does not cover the long bar which may be rotating differentially with respect to the bar/bulge. In Section~\ref{Section::PatternSpeed} we derive new expressions for estimating the pattern speed from proper motion data using the continuity equation. In Section~\ref{Section::Application} we apply these expressions to the bar/bulge data. In Section~\ref{Section::Recovery} we demonstrate how well the method works when applied to a simulation of a dynamically-formed bar/bulge and test the limitations and assumptions of the presented method. In Appendix~\ref{Appendix:z0} we generalise the expressions for the pattern speed to the account for the Sun's small offset from the Galactic plane, and in Appendix~\ref{appendix:lf} we validate our choice of luminosity function using local Gaia data.

\section{The pattern speed of the bar}\label{Section::PatternSpeed}
\cite{TremaineWeinberg} introduced a method for measuring the pattern speed of a barred disc galaxy using only the continuity equation and the assumption that the pattern is steady.  \cite{KuijkenTremaine1991} and \cite{Debattista2002} adapted the formalism for use with line-of-sight velocities in the Milky Way with \cite{Debattista2002} applying the formulae to OH/IR stars across the Galactic disc. Whilst \cite{TremaineWeinberg} and \cite{KuijkenTremaine1991} worked in 2D, we shall follow \cite{Debattista2002} who provided expressions for 3D density distributions (but only in the case of line-of-sight velocities). 

We assume the stars follow a tracer density $\rho(\bs{x},t)$ rotating at a constant pattern speed $\Omega_\mathrm{p}$. In terms of the Cartesian coordinates $(x,y,z)$ in the non-rotating disc frame, the continuity equation is given by
\begin{equation}
    \frac{\partial\rho}{\partial t} +\nabla\cdot(\rho\bs{v}) =\Omega_\mathrm{p}\Big[y\frac{\partial\rho}{\partial x}-x\frac{\partial\rho}{\partial y}\Big]+\nabla\cdot(\rho\bs{v})=0.
\label{eqn::continuity}
\end{equation}

Introducing the standard Galactic coordinates $(\ell,b)$ along with distance $s$ and corresponding transverse and line-of-sight velocities $v_\ell$, $v_b$ and $v_{||}$, the continuity equation becomes
\begin{equation}
\begin{split}
    &\Omega_\mathrm{p}\Big[-R_0\cos b\sin\ell\frac{\partial\rho}{\partial s}+\Big(1-\frac{R_0\cos\ell}{s\cos b}\Big)\frac{\partial\rho}{\partial\ell}
    +\frac{R_0}{s}\sin\ell\sin b\frac{\partial\rho}{\partial b}\Big]\\
    &+\frac{1}{s^2}\frac{\partial(s^2\rho v_{||})}{\partial s}+\frac{1}{s\cos b}\frac{\partial(\rho v_\ell)}{\partial \ell}+\frac{1}{s\cos b}\frac{\partial(\rho v_b \cos b)}{\partial b}=0.
\end{split}
\end{equation}
$R_0$ is the distance to the Galactic centre and the choice of coordinate systems is such that $\Omega_\mathrm{p}>0$ for the Galactic bar (i.e. a left-handed $(x,y,z)$ system). Here and in the subsequent expressions we have assumed the Sun is in the Galactic plane ($z=0$). In Appendix~\ref{Appendix:z0} we give the expressions incorporating the small offset ($\sim25\,\mathrm{pc}$) due to the Sun's height above the Galactic plane. Mirroring the method of \cite{TremaineWeinberg}, we multiply by $s^2\cos b$ and integrate with respect to $s$ from $0$ to $\infty$. We use the fact that $s^2\rho$ vanishes at $0$ and $\infty$ (provided $\rho$ falls off faster than $s^2$ at $\infty$ -- reasonable for a barred tracer) and integrate once by parts to write
\begin{equation}
\begin{split}
    &\Omega_\mathrm{p}\Big[2 R_0\cos^2 b\sin\ell\int_0^\infty\mathrm{d}s\,s\rho
    +\int_0^\infty\mathrm{d}s\,(s^2\cos b-
    R_0 s\cos\ell)\frac{\partial\rho}{\partial\ell}
    \\&
    +R_0\sin\ell\sin b\cos b\int_0^\infty\mathrm{d}s\,s\frac{\partial\rho}{\partial b}\Big]\\
    &+\int_0^\infty\mathrm{d}s\,s\frac{\partial(\rho v_\ell)}{\partial \ell}+\int_0^\infty\mathrm{d}s\,s\frac{\partial(\rho v_b \cos b)}{\partial b}=0.
\end{split}
\end{equation}
Noting $\partial/\partial b(\sin b\cos b)=2\cos^2b-1$, we can combine the $\ell$ and $b$-derivatives with the first integral such that
\begin{equation}
\begin{split}
    &\Omega_\mathrm{p}R_0\sin\ell\frac{\partial}{\partial b}\int_0^\infty\mathrm{d}s\,\rho s \sin b\cos b
     \\&+\Omega_\mathrm{p}\frac{\partial}{\partial\ell}\int_0^\infty\mathrm{d}s\,
     \rho s (s\cos b-
    R_0\cos\ell)
    \\
    &+\int_0^\infty\mathrm{d}s\,s\frac{\partial(\rho v_\ell)}{\partial \ell}+\int_0^\infty\mathrm{d}s\,s\frac{\partial(\rho v_b \cos b)}{\partial b}=0.
\end{split}
\label{eqn::lbderiv}
\end{equation}
We proceed by integrating in $b$ from $-\pi/2$ to $\pi/2$ imposing the condition that $\rho(b=\pm\pi/2)=0$ for all $s$ (this is valid provided the solar radius $R_0$ encompasses the entirety of the non-axisymmetry). This leaves only the second and third terms. A final integration in $\ell$ from $-\pi$ to $\ell$ (again requiring the non-axisymmetry to be completely encompassed such that $\rho(\ell=\pi)=0$ for all $s$) reduces the expression to 
\begin{equation}
\begin{split}
    \Omega_\mathrm{p}(\ell) &= \frac{\int_{-\pi/2}^{\pi/2}\mathrm{d}b\int_0^\infty\mathrm{d}s\,s\rho v_\ell}{\int_{-\pi/2}^{\pi/2}\mathrm{d}b\int_0^\infty\mathrm{d}s\,
    s\rho(R_0 \cos\ell-s\cos b)}\\&\equiv\frac{\langle v_\ell \rangle_{b,s}}{\langle R_0\cos\ell-s\cos b\rangle_{b,s}},
    \end{split}
\end{equation}
where we have introduced the notation $\langle\rangle_{i,j}$ to denote multiplying by $s\rho$ and averaging over $i$ and $j$. Here we have written $\Omega_\mathrm{p}$ as a function of $\ell$ to make clear that the right-hand side is a function of $\ell$. However, provided the assumptions are satisfied, $\Omega_\mathrm{p}$ is a constant. Note we are free to multiply numerator and denominator by a general function $f(\ell)$ and integrate over $\ell$. This is the equivalent of \cite{TremaineWeinberg}'s $h(Y)$ which they recommend to be odd to minimise the effect of centering errors. For our application, it may be desirable to make $f(\ell)$ the inverse of the variance of the result for each $\ell$. We will see how we can combine the estimates probabilistically.

We observe that the derived expression is independent of the current bar angle $\alpha$. For a bar with no azimuthal extent oriented at $\alpha$, $\rho=|R|^{-1}\rho_R(|R|)\delta(\phi-\alpha)\rho_z(z)$, the Galactic longitude velocity $v_\ell(\ell)=R\Omega_\mathrm{p}\cos(\ell+\alpha)$ where $R$ is the cylindrical polar distance from the centre of the Galaxy to the intercept between the bar and the line-of-sight (taking negative sign for $\ell<0$). Therefore, $R\cos(\ell+\alpha)=R\cos\ell\cos\alpha-R\sin\ell\sin\alpha=R_0\cos\ell-s\cos b$ as $R_0=s\cos b\cos\ell+R\cos\alpha$ and $s\cos b\sin\ell=R\sin\alpha$. We recover the expected result.

We can return to equation~\eqref{eqn::lbderiv} and proceed in a similar fashion. We integrate instead in $\ell$ from $-\pi$ to $\pi$ and $b$ from $-\pi/2$ to $b$ to find
\begin{equation}
        \Omega_\mathrm{p}(b) = -\frac{\langle v_b\rangle_{\ell,s}}{
    R_0\sin b\langle \sin \ell\rangle_{\ell,s}}.
\end{equation}
The $\sin b$ in the denominator naturally makes this estimator poorly suited to measuring $\Omega_\mathrm{p}$ as any noise in $v_b$ is amplified for small $b$. The equivalent expression for the line-of-sight velocities is obtained by multiplying the continuity equation by $\cos b$, integrating over all $\ell$ and $b$ (integrating by parts once) and integrating $s$ from $\infty$ to $s$:
\begin{equation}
    \Omega_\mathrm{p}(s) = \frac{\langle v_{||}\cos b\rangle_{\ell,b}}{
    R_0\langle \sin\ell\cos^2 b\rangle_{\ell,b}}.
\end{equation}
This expression is given by \cite{Debattista2002} who proceed to multiply both numerator and denominator by a general function $f(s)$ and integrate over all distance, $s$. Note that our three estimators have all reduced the 3D continuity equation to considering flows in a single dimension by integrating over the other two coordinates. Each expression in essence measures the rotational velocity divided by the distance to the Galactic centre. However, they are not exactly equivalent. This is because a bar is preferentially supported by prograde orbits. As \cite{Qin2015} showed, simply using the rotational velocity at a few spatial locations to estimate the pattern speed leads to biases. The geometric factors and averaging in the estimators are a necessary component.

These estimators are valid provided (i) the pattern is steady, (ii) the integration encompasses the entirety of the non-axisymmetry and (iii) there is a single pattern speed. It is likely that the pattern speed of the bar is evolving slowly -- the impact of this can be tested using our reference simulation. The validity of the second and third assumptions is less clear, particularly in the presence of other non-axisymmetries such as spiral arms. Our modelling extracts non-axisymmetries only in the red giant populations which do not trace recent star formation and so are likely free from small-scale non-axisymmetries. It is likely that the entire bar rotates at a single pattern speed as differentially rotating triaxial structures rapidly exchange angular momentum.

\subsection{Using heliocentric velocities}

Our expressions were derived using velocities measured in the Galactic rest frame. For heliocentric velocities (denoted by primes), we must first correct for the solar reflex motion as
\begin{equation}
    \begin{split}
        v_{||}&=v'_{||}+u_\odot\cos\ell\cos b+v_\odot\sin\ell\cos b+w_\odot\sin b,\\
        v_{\ell}&=v'_{\ell}-u_\odot\sin\ell+v_\odot\cos\ell,\\
        v_{b}&=v'_{b}-u_\odot\cos\ell\sin b-v_\odot\sin\ell\sin b+w_\odot\cos b,
    \end{split}
\end{equation}
where $(u_\odot,v_\odot,w_\odot)$ is the solar velocity in the Galactic rest frame (positive $u$ towards the centre of the Galaxy and positive $v$ in the direction of Galactic rotation). Our pattern speed expressions in terms of heliocentric coordinates are then given by
\begin{equation}
    \Omega_\mathrm{p}(s)=\frac{v_\odot}{R_0}+ \frac{ \langle(v'_{||}+u_\odot\cos\ell\cos b+w_\odot\sin b)\cos b\rangle_{\ell,b}}{\langle
    R_0 \sin\ell\cos^2 b\rangle_{\ell,b}}.
\end{equation}
\begin{equation}
    \Omega_\mathrm{p}(b)=\frac{v_\odot}{R_0}- \frac{\langle v'_{b}-u_\odot\cos\ell\sin b+w_\odot\cos b\rangle_{\ell,s}}{\langle
    R_0\sin\ell\sin b\rangle_{\ell,s}}.
\end{equation}
\begin{equation}
    \Omega_\mathrm{p}(\ell)=\frac{v_\odot-u_\odot\tan\ell+\mathcal{F}}{R_0-\mathcal{K}}
    ,\>
    \mathcal{F}=\frac{\langle v'_\ell\rangle_{b,s}}{\langle \cos\ell\rangle_{b,s}},\>\mathcal{K}=\frac{\langle s\cos b\rangle_{b,s}}{\langle\cos\ell\rangle_{b,s}}
    \label{eqn::fk_equation}
\end{equation}
Note that for $\Omega_\mathrm{p}(s)$ and $\Omega_\mathrm{p}(b)$, the pattern speed is completely degenerate with $v_\odot/R_0$ for all $\ell,b,s$, whilst for $\Omega_\mathrm{p}(\ell)$ the degeneracy is more complex and depends on the geometric factor $\mathcal{K}$. In the thin bar limit, $\mathcal{K}=R_0$ when the viewing angle is orthogonal to the bar as in this case $v_\ell$ contains no contribution from the azimuthal rotation. 

The impact of the solar velocity depends on the estimator employed. $u_\odot$ and $w_\odot$ are well measured as $u_\odot=(11.1\pm0.7\pm1\,\mathrm{sys}.)\,\mathrm{km\,s^{-1}}$ and $w_\odot=(7.25\pm0.36\pm0.50\,\mathrm{sys}.)\,\mathrm{km\,s^{-1}}$ \citep[][although the local standard of rest may have a non-zero $u$ velocity]{Schoenrich2010}. Uncertainties in $u_\odot$ primarily affect $\Omega_\mathrm{p}(s)$ as $\Omega_\mathrm{p}(\ell)$ contains $u_\odot\tan\ell$ (and $|\ell|<10\,\mathrm{deg}$). Assuming Sgr A* is a rest with respect to the bulge, $v_\odot/R_0$ is well constrained from the proper motion of Sgr A* as $\mu_{\ell,A*}=-v_\odot/(4.74 R_0)=(-6.379\pm0.026)\,\mathrm{mas\,yr}^{-1}$ \citep{Reid2004}. Therefore, we conclude that the uncertainty in the pattern speed estimators, $\Omega_\mathrm{p}(s)$ and $\Omega_\mathrm{p}(b)$, comes primarily from the unknown Galactic centre distance $R_0$. Fixing $\mu_{\ell,A*}$ and ignoring the impact of $u_\odot$ and $w_\odot$, we have that both $\Omega_\mathrm{p}(s)$ and $\Omega_\mathrm{p}(b)$ fall with $R_0$ like $\Omega_\mathrm{p}=k_1+k_2/R_0$ whilst for $\Omega_\mathrm{p}(\ell)$ we find
\begin{equation}
    \Omega_\mathrm{p}(\ell)=\frac{-4.74\mu_{\ell,A*}R_0+\mathcal{F}}{R_0-\mathcal{K}}.
\end{equation}
This more complex behaviour allows the possibility of measuring $R_0$ and $\Omega_\mathrm{p}$ from the data. However, with $\mu_{\ell,A*}$ fixed, inference of $R_0$ is degenerate with a fractional distance systematic (i.e. a constant absolute magnitude offset). The combination of both $\Omega_\mathrm{p}(s)$ and $\Omega_\mathrm{p}(\ell)$ puts stronger restrictions on $R_0$, $v_\odot$, $u_\odot$ and $\Omega_\mathrm{p}$, potentially testing the assumption of steady state. However, such an analysis would require a spectroscopic dataset distributed over all $b$ and $\ell$.

\subsection{Estimators from Jeans' equations}

We briefly discuss the possibility of using higher order equations derived from the collisionless Boltzmann equation (CBE) to measure the pattern speed. These could provide tighter constraints on the pattern speed. After the continuity equation, the next two equations (obtained by multiplying the CBE by $v_j$ and $v_jv_k$ respectively and integrating over all velocities) are
\begin{equation}
\begin{split}
\frac{\partial(\rho\overline{v}_j)}{\partial t}+\frac{\partial(\rho\overline{v_iv_j})}{\partial x_i}+\rho\frac{\partial\Phi}{\partial x_j}&=0,\\
\frac{\partial(\rho\overline{v_jv_k})}{\partial t}+\frac{\partial(\rho\overline{v_iv_jv_k})}{\partial x_i}+\overline{v}_j\rho\frac{\partial\Phi}{\partial x_k}+\overline{v}_k\rho\frac{\partial\Phi}{\partial x_j}&=0,
\end{split}
\label{eqn::higherorder}
\end{equation}
where we have introduced the potential $\Phi$. Using the first of these equations combined with the continuity equation and integrating over all space \citep[c.f.][]{BinneyTremaine2008}, we find
the tensor-virial theorem in the form \citep{ChandrasekharEFE}
\begin{equation}
2K_{ij}+W_{ij}+\Omega_\mathrm{p}^2(J_{ij}-\delta_{3i}J_{3j})+2\levicivita{i}{k}{3}\Omega_\mathrm{p}\int_V\mathrm{d}^3\bs{x}\,\rho w_k x_j=0,
\end{equation}
where the velocities are in the rotating frame $w_i=v_i-\levicivita{i}{3}{k}\Omega_\mathrm{p}x_k$ and the tensors are
\begin{equation}
    K_{ij}=\int_V\mathrm{d}^3\bs{x}\rho w_i w_j,\>\>
    W_{ij}=\int_V\mathrm{d}^3\bs{x}\rho x_i a_j,\>\>
    J_{ij}=\int_V\mathrm{d}^3\bs{x}\rho x_i x_j.
\end{equation}
Imposing triaxial symmetry removes the final Coriolis-like term leaving a balance between random kinetic energy, potential energy and the rotational energy of the figure. Also, all off-diagonal relations are zero. We are then left with three relations between the pattern speed and the three unknown potential energy tensor diagonal components. We therefore cannot use these relations to infer the pattern speed without  more knowledge of the potential (e.g., that the density is ellipsoidally stratified).

Another approach is to combine the two expressions in equation~\eqref{eqn::higherorder} to eliminate the potential derivatives \citep{KuijkenTremaine1991}
\begin{equation}
\frac{\partial\sigma_{jk}^2}{\partial t}+\sigma^2_{ik}\frac{\partial\overline{v}_j}{\partial x_i}+\sigma^2_{ij}\frac{\partial\overline{v}_k}{\partial x_i}+\overline{v}_i\frac{\partial\sigma^2_{jk}}{\partial x_i}
+\frac{1}{\rho}\frac{\partial(\rho\sigma^3_{ijk})}{\partial x_i}=0,
\end{equation}
where $\sigma^2_{ij}\equiv\overline{(v_i-\overline{v_i})(v_j-\overline{v_j})}$ and $\sigma^3_{ijk}\equiv\overline{(v_i-\overline{v_i})(v_j-\overline{v_j})(v_k-\overline{v_k})}$. If we neglect the third order moments $\sigma_{ijk}^3$ then this equation gives a simple conservation law for the kinetic pressure. Assuming a constant pattern speed, we can perform the same replacement giving
\begin{equation}
\Omega_\mathrm{p}\Big[y\frac{\partial\sigma^2_{jk}}{\partial x}-x\frac{\partial\sigma^2_{jk}}{\partial y}\Big]+\sigma^2_{ik}\frac{\partial\overline{v}_j}{\partial x_i}+\sigma^2_{ij}\frac{\partial\overline{v}_k}{\partial x_i}+\overline{v}_i\frac{\partial\sigma^2_{jk}}{\partial x_i}=0.
\end{equation}
However, here we observe that the right-hand side involves all three velocity components and all three components of $\sigma^2_{ij}$ (for fixed $j$). This is not a limitation in the Tremaine-Weinberg method where terms involving unknown velocities are integrated out. However, we cannot write the terms on the right-hand side of this equation purely as sums of derivatives making replication of the Tremaine-Weinberg method hard. With a full knowledge of the velocity field, this equation (with or without the inclusion of the $\sigma_{ijk}^3$ term) offers an attractive way to measure the pattern speed from the dispersion (and mean velocity) field without knowledge of the potential.

\subsection{Probablistic approach}

We develop a framework for the application of the estimator $\Omega_\mathrm{p}(\ell)$ provided in the previous section that accounts for uncertainty in the observables as well as the Galactic parameters. Such an approach is necessary for a robust estimate of $\Omega_\mathrm{p}$ with associated uncertainties. The problem is essentially equivalent to a linear regression with correlated uncertainties in both `$x$' and `$y$' \citep[e.g.][]{Me95} and a prior on the intercept.

We consider as our measurements at each $\ell$ the `vector' $\bs{\mathcal{X}}_i=(\mathcal{F},\mathcal{K})_i$ with corresponding covariance $\bs{\Sigma}_{\mathcal{X}i}$. Both the numerator and denominator of $\Omega_\mathrm{p}(\ell)$ tend to zero for $\ell\rightarrow0$ i.e. the centre of the bar/bulge is not translating with respect to the Galaxy and the centre is located at $R_0$. Therefore, both numerator and denominator can be modelled by a power series with no constant coefficient:
\begin{equation}
f(\ell)=v_\odot-u_\odot\tan\ell+\mathcal{F}=\Omega_\mathrm{p}(R_0-\mathcal{K})=\Omega_\mathrm{p}R_0\sum^{n=N_\mathrm{max}}_{n=1}c_n(\tan\ell)^n.
\end{equation}
We expand in terms of $\tan\ell$ as for a needle-thin bar at angle $\alpha$ $R_0-\mathcal{K}=R_0\tan\ell/\tan(\ell+\alpha)$. Inspecting our reference simulation (see Section~\ref{Section::Recovery}), we find that $1\lesssim N_\mathrm{max}\lesssim3$ is appropriate, so we set $N_\mathrm{max}=3$ for application to the data. This approach is chosen rather than modelling $\Omega_\mathrm{p}(\ell)$ directly as the uncertainty in $\Omega_\mathrm{p}(\ell)$ becomes larger as $\ell\rightarrow0$ whilst $\mathcal{F}$ and $\mathcal{K}$ are well behaved.

We construct the likelihood
\begin{equation}
\begin{split}
    \bs{\mathcal{X}}_i&\sim\mathcal{N}(\bs{f}(\ell_i),\bs{\Sigma}_{\mathcal{X}i}+\bs{\Sigma}_f),\\
    \bs{f}(\ell_i)&=\lambda\Big( f(\ell_i)-v_\odot+u_\odot\tan\ell_i,R_0-\frac{ f(\ell_i)}{\Omega_\mathrm{p}}\Big),\\ \bs{\Sigma}_f&=\mathrm{diag}(\sigma^2_\mathcal{F},\sigma^2_\mathcal{K}),
\end{split}
\label{Eqn::OmegaModel}
\end{equation}
where $\mathcal{N}(\bs{\mu},\bs{\Sigma})$ is a multidimensional normal distribution with mean $\bs{\mu}$ and covariance matrix $\bs{\Sigma}$. The diagonal covariance matrix $\bs{\Sigma}_f$ represents an intrinsic scatter in $\mathcal{F}$ and $\mathcal{K}$ which could arise from the neglected boundary terms. $\lambda$ is a fractional distance systematic, so all observed distances are a factor $\lambda$ too large. We adopt the following priors:
\begin{equation}
    \begin{split}
        R_0/\,\mathrm{kpc}&\sim\mathcal{N}(\mu_{R0},\sigma_{R0}),\\
        \mu_{\ell,A*}/\,\mathrm{mas\,yr}^{-1}&\sim\mathcal{N}(-6.379,0.026),\\
        \lambda&\sim\mathcal{N}(1,0.05),\\
        u/\,\mathrm{km\,s}^{-1}&\sim\mathcal{N}(11.1,1.2),\\
        \ln\sigma_\mathcal{F}/\mathrm{km\,s}^{-1}&\sim\mathcal{U}(-10,10),\\
        \ln\sigma_\mathcal{K}/\mathrm{kpc}&\sim\mathcal{U}(-10,10),
    \end{split}
\end{equation}
where $\mathcal{N}(\mu,\sigma)$ is a normal with mean $\mu$ and standard deviation $\sigma$ and $\mathcal{U}(a,b)$ is uniform between $a$ and $b$.
For $R_0$ we consider three priors: $\mathcal{N}(8.12\,\mathrm{kpc},0.03\,\mathrm{kpc})$ from \cite{GravityCollab2018}, $\mathcal{N}(8.2\,\mathrm{kpc},0.09\,\mathrm{kpc})$ from \cite{McMillan2017} (where it should be noted that \cite{McMillan2017} uses the proper motion of Sgr A* in the estimate of $R_0$ leading to a covariance between $v_\odot$ and $R_0$) and $\mathcal{N}(8.12\,\mathrm{kpc},2\,\mathrm{kpc})$. The second prior is approximately that suggested by \cite{BlandHawthornGerhard2016} from combining estimates from multiple studies. The third prior is an uninformative prior. We also consider an uninformative prior on $\mu_{\ell,A*}$ of $\mathcal{N}(-6.379\,\mathrm{mas\,yr}^{-1},1\,\mathrm{mas\,yr}^{-1})$. Our prior on $\lambda$ corresponds approximately to considering magnitude systematics of $\sim0.1\,\mathrm{mag}$. We write the model in Stan and sample using the NUTS sampler \citep{Hoffman2011}.

\section{Application to data}\label{Section::Application}

We apply the derived framework to the data produced in Paper I. In that work, we derived the mean transverse velocities $v_\ell$ and $v_b$ and corresponding uncertainties as a function of distance across the bar/bulge region from $\sim45$ million giant stars in VVV. We used a combination of proper motions from Gaia DR2 \citep{Gaia2018} and VIRAC v1.1 \citep{Sm18}, an astrometric catalogue derived from the VVV observations calibrated absolutely using Gaia DR2. In small fields of $0.2\,\mathrm{deg}\times0.2\,\mathrm{deg}$ across the VVV bulge footprint, the bulge giants (both red giant branch and red clump) were selected in a unextincted colour-magnitude box of $0.4<(J-K_s)_0<1$ and $11.5<K_{s0}<14.5$ using a two-dimensional extinction map from the method of \cite{Gonzalez2011}. First, the density in each field was measured assuming a luminosity function \citep[taken from][]{Simion2017} and accounting for incompleteness, and then the density distribution was used to extract the first two moments of the transverse velocity distributions as a function of distance by probabilistically considering the possible velocities for each star given its $K_{s0}$ magnitude. Here we utilise the mean longitudinal transverse velocities with location $v_\ell(\ell,b,s)$ and the density field $\rho(\ell,b,s)$ computed on a grid in log(distance), $\ell$ and $b$. The integrals are computed using the trapezoidal rule but using Simpson's rule instead gives very similar results. We fill in the region $5<b/\,\mathrm{deg}<10$ using $-10<b/\,\mathrm{deg}<-5$ assuming reflection symmetry in $b=0$. This increases the estimates of the pattern speed by $\sim2\,\mathrm{km\,s}^{-1}\mathrm{kpc}^{-1}$. The uncertainties $\bs{\Sigma}_{\bs{\mathcal{X}}_i}$ are computed through propagation of uncertainties in density and velocity. We recall that due to the procedure in Paper I the uncertainties in the velocities are probably underestimated as they only account for proper motion uncertainties not uncertainties in the density (or luminosity function) which propagate to increased uncertainties in the physical velocities.

In Table~\ref{Table::results}, we show the results of applying the method to different subsets of the data and the different priors. We report the estimates of $R_0$, $\mu_{\ell,A*}$ and $\Omega_\mathrm{p}$ as well as the chi-squared per datapoint which is given by
\begin{equation}
    \chi^2/N \equiv \frac{1}{N}\sum^N_i\frac{(\Omega_\mathrm{p}-\Omega_\mathrm{p}(\ell_i))^2}{\sigma_\Omega^2(\ell_i)},
\end{equation}
where $\Omega_\mathrm{p}$ is the median from the MCMC chain, $\Omega_\mathrm{p}(\ell_i)$ the estimate for bin $i$ and $\sigma_\Omega(\ell_i)$ the associated uncertainty (from propagating the uncertainties in $\mathcal{F}$ and $\mathcal{K}$). 

We see that when using all of the data with the `tight' prior we obtain $\Omega_\mathrm{p}=(37.2\pm3.3)\,\mathrm{km\,s}^{-1}\mathrm{kpc}^{-1}$, but the $\chi^2/N=19.7$ is poor and $\mu_{\ell,A*}$ is $2\sigma$ away from the prior. Using the 'loose' prior, we find the pattern speed remains similar, the $\chi^2/N=12.3$ reduces but still $\mu_{\ell,A*}$ is in tension. Relaxing the $R_0$ prior further only weakly improves $\chi^2/N$ but $R_0$ does not stray significantly. Using a weak prior on $\mu_{\ell,A*}$, we see the data `wants' to reduce $|\mu_{\ell,A*}|$ producing a low pattern speed of $\Omega_\mathrm{p}=(23.7\pm2.7)\,\mathrm{km\,s}^{-1}$ but not with a significantly improved $\chi^2/N$. 

We next have tried separating the data into $\ell>0$ and $\ell<0$. The two models with lowest $\chi^2/N$ are $\ell<0$ `No $R_0$' with  $\Omega_\mathrm{p}=(34.7\pm5)\,\mathrm{km\,s}^{-1}$ but producing an inconsistent $R_0$ of $(9.42\pm0.45)\,\mathrm{kpc}$ and $\ell>0$ `No $R_0$' with $\Omega_\mathrm{p}=(30.9\pm1.2)\,\mathrm{km\,s}^{-1}$ but producing an inconsistent $\mu_{\ell,A*}$ of $(-6.24\pm0.01)\,\mathrm{mas\,yr}^{-1}$. We next tried removing the central regions $|\ell|>2.5\,\mathrm{deg}$ as these produce the noisiest estimates of $\Omega_\mathrm{p}$ as $v_\ell$ is small. The set of models for $\ell>2.5\,\mathrm{deg}$ produce very satisfactory results with low $\chi^2/N\approx1.3-1.5$ and both $R_0$ and $\mu_{\ell,A*}$ consistent with expectations both with and without the prior. All models give $\Omega_\mathrm{p}\approx(41\pm3)\,\mathrm{km\,s}^{-1}\mathrm{kpc}^{-1}$. For $\ell<-2.5\,\mathrm{deg}$ the situation is less satisfactory with higher $\chi^2/N$. The tension is illustrated by the `No $R_0$' model which produces a much too high estimate of $R_0$ but $\Omega_\mathrm{p}$ consistent with $\ell>2.5\,\mathrm{deg}$. When combining both positive and negative $\ell$, we obtain $\Omega_\mathrm{p}\approx(30\pm1)\,\mathrm{km\,s}^{-1}\mathrm{kpc}^{-1}$ but the $\chi^2/N$ are higher and we always recover $\mu_{\ell,A*}$ many $\sigma$ away from the measured value.

\begin{figure}
    \centering
    \includegraphics[width=\columnwidth]{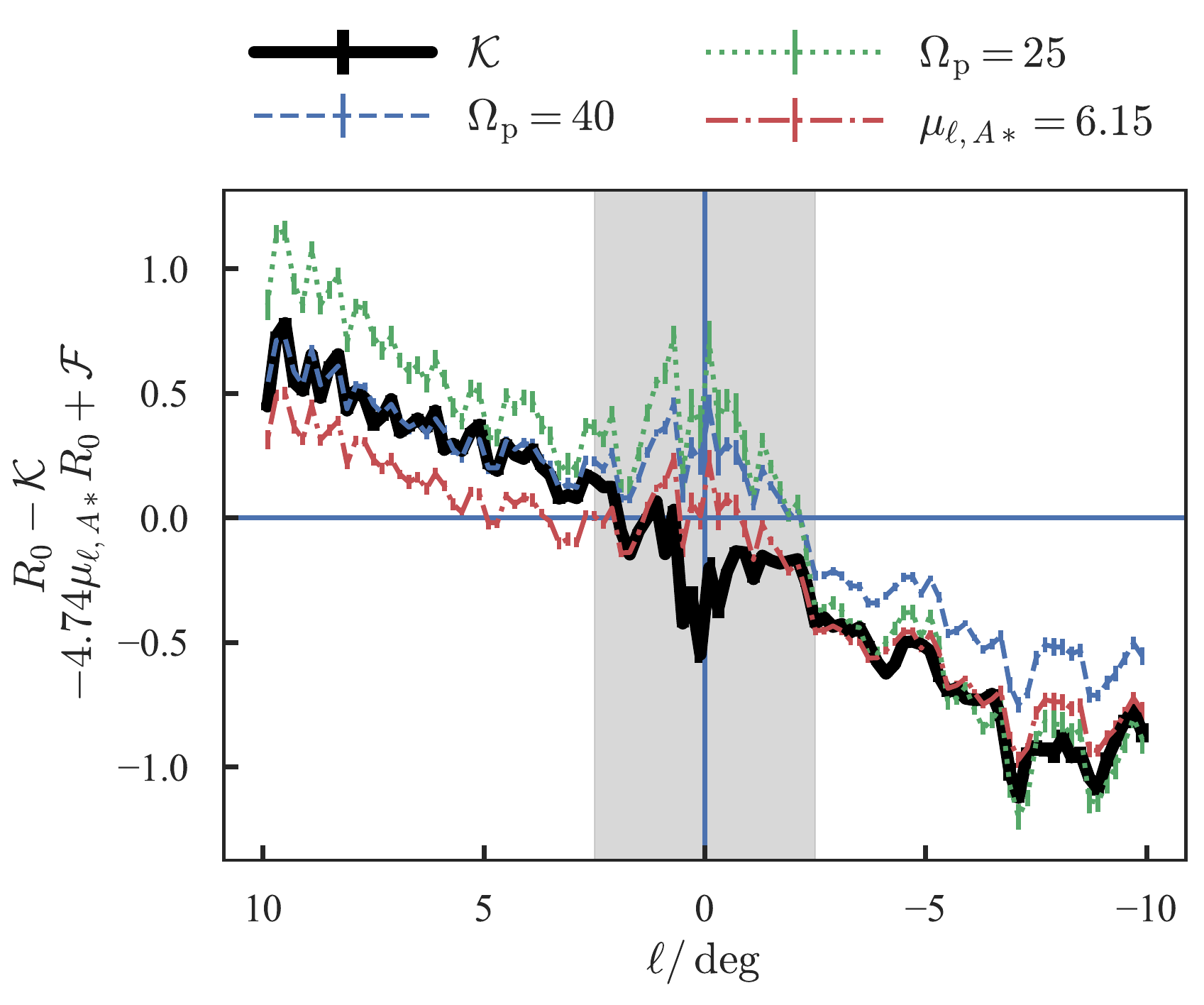}
    \caption{Terms in our pattern speed estimator $(R_0-\mathcal{K})$ (black using $R_0=8.12\,\mathrm{kpc}$) and $(-4.74\mu_{\ell,A*}R_0-u_\odot\tan\ell+\mathcal{F})/\Omega_\mathrm{p}$ (coloured) in units of $\mathrm{kpc}$ and binned in $\ell$. When the two terms are identical, the pattern speed is constant and correct across the bar. The blue dashed line uses $\mu_{\ell,A*}=-6.379\,\mathrm{mas\,yr}^{-1}$ and $\Omega_\mathrm{p}=40\,\mathrm{km\,s}^{-1}\mathrm{kpc}^{-1}$, green dotted reduces $\Omega_\mathrm{p}=25\,\mathrm{km\,s}^{-1}\mathrm{kpc}^{-1}$ and red dash-dot increases $\mu_{\ell,A*}=-6.15\,\mathrm{mas\,yr}^{-1}$. We see the far-side of the bar requires a lower pattern speed or higher $\mu_{\ell,A*}$. Within the shaded region $|\ell|<2.5\,\mathrm{deg}$ the estimator is noisy.}
    \label{fig:fk_balance}
\end{figure}

We can understand these results by inspecting Fig.~\ref{fig:fk_balance} which shows the two terms in equation~(\ref{eqn::fk_equation}) that must be equal for the pattern speed to be constant across the bar. We see a pattern speed of $\Omega_\mathrm{p}=40\,\mathrm{km\,s}^{-1}\mathrm{kpc}^{-1}$ produces a consistent result for $\ell>0$ but the magnitude of the velocities is too small (or the distances too small) for $\ell<0$. This is fixed by adopting $\Omega_\mathrm{p}=25\,\mathrm{km\,s}^{-1}\mathrm{kpc}^{-1}$ or $\mu_{\ell,A*}=-6.15\,\mathrm{mas\,yr}^{-1}$ (many $\sigma$ from the measured value) but produces a poorer fit for $\ell>0$. Note the grey region $|\ell|<2.5\,\mathrm{deg}$ where the estimator is noisy. 

\begin{figure}
    \centering
    \includegraphics[width=\columnwidth]{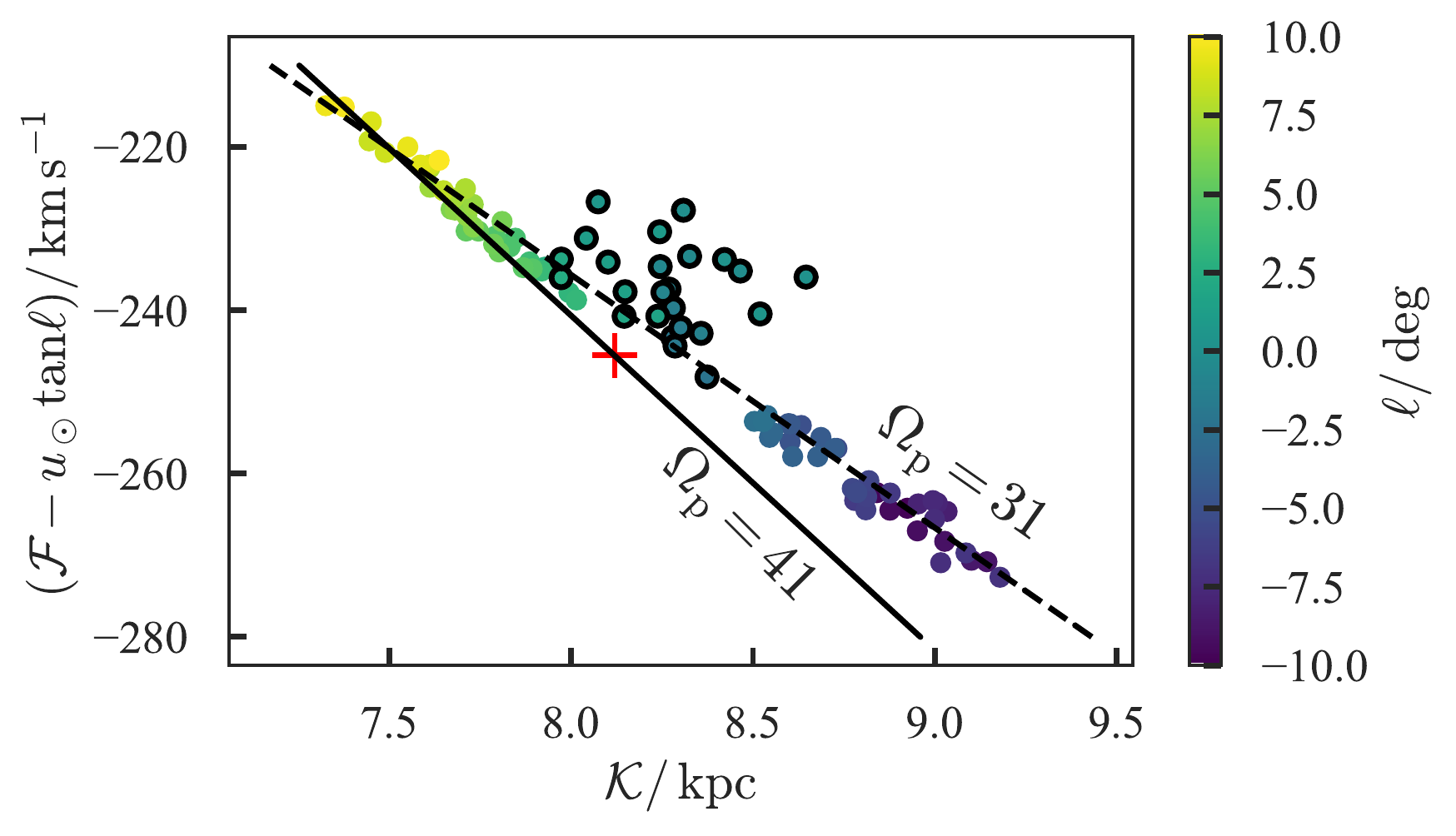}
    \caption{Terms in our pattern speed estimator $\mathcal{K}$ vs. $-u_\odot\tan\ell+\mathcal{F}$ for each $\ell$ bin (given by the colour). The gradient of this plot corresponds to the pattern speed. The points with black edges have $|\ell|<2.5\,\mathrm{deg}$ and clearly lie away from the trend. The red cross shows the Galactic centre (at $R_0=8.12\,\mathrm{kpc}$ and $-4.74\mu_{\ell,A*}R_0=245.5\,\mathrm{km\,s}^{-1}$). The black solid line shows approximately our best fit to $\ell>2.5\,\mathrm{deg}$ with $\Omega_\mathrm{p}=41\,\mathrm{km\,s}^{-1}\mathrm{kpc}^{-1}$ whilst the dashed line shows the best fit to the entire data with $\Omega_\mathrm{p}=31\,\mathrm{km\,s}^{-1}\mathrm{kpc}^{-1}$ which doesn't pass through the Galactic centre.}
    \label{fig:fk_gradient}
\end{figure}

An alternative way of viewing the data is plotting $\mathcal{F}-u_\odot\tan\ell$ against $\mathcal{K}$ as shown in Fig.~\ref{fig:fk_gradient}. The gradient in this plane gives the pattern speed. We observe how the points with $|\ell|<2.5\,\mathrm{deg}$ are significantly deviant any straight line fit validating their removal. We see that the near-side of the bar produces a linear fit with gradient $\Omega_\mathrm{p}=41\,\mathrm{km\,s}^{-1}\mathrm{kpc}^{-1}$ and intercept consistent with the Galactic centre position and velocity. However, inclusion of the far-side reduces the gradient to $\Omega_\mathrm{p}=31\,\mathrm{km\,s}^{-1}\mathrm{kpc}^{-1}$ but produces an inconsistent intercept.

\begin{figure}
    \centering
    \includegraphics[width=\columnwidth]{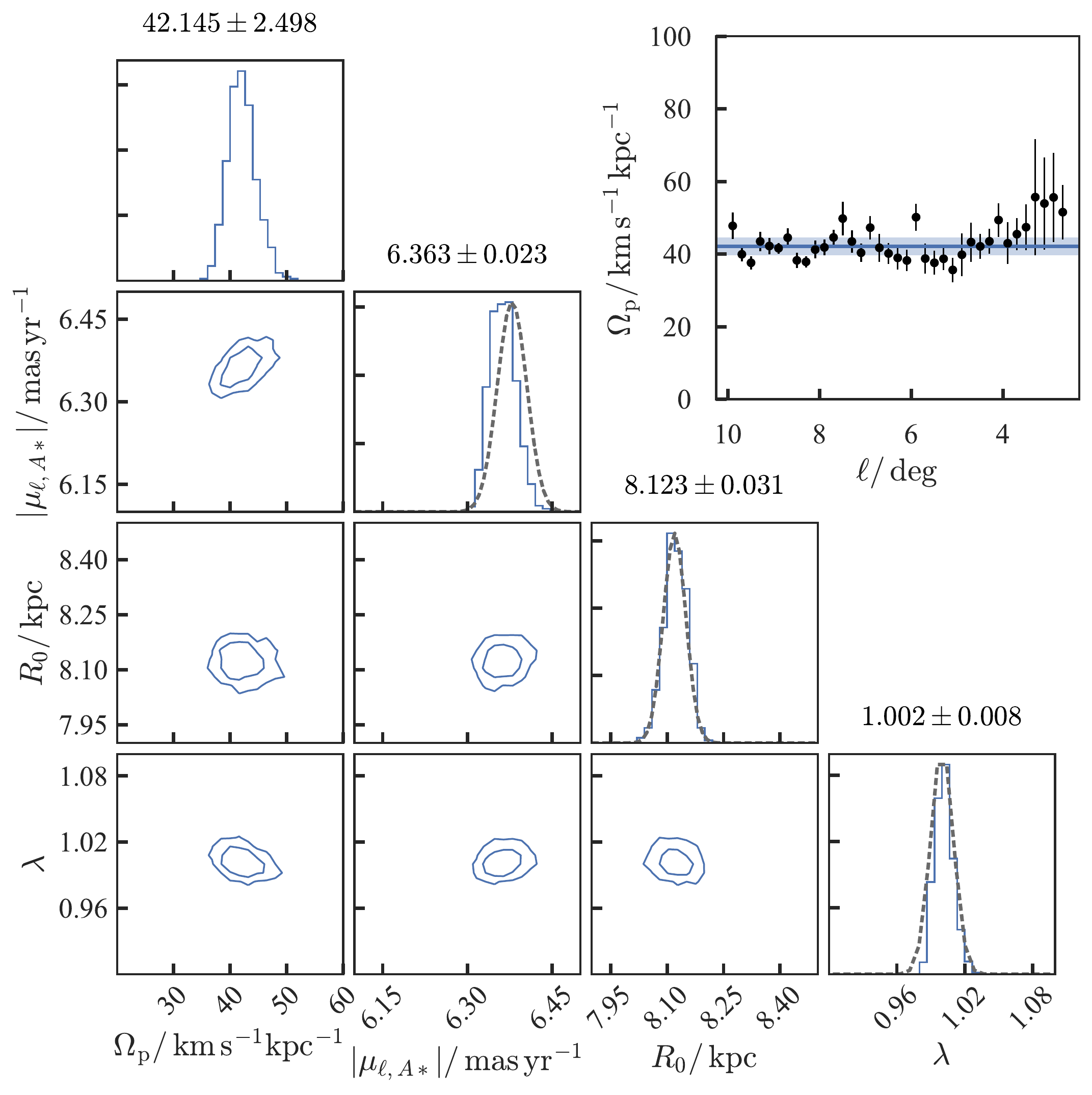}
    \caption{Posterior distributions for the model parameters when applied to data with $\ell>2.5\,\mathrm{deg}$. The grey dotted lines show the assumed prior. The top right inset shows the pattern speed estimate at each $\ell$ along with the median and $1\sigma$ error band for the pattern speed. $\lambda$ is a fractional distance systematic.}
    \label{fig:data_application}
\end{figure}

In Fig.~\ref{fig:data_application}, we show the results from our most successful model $\ell>2.5\,\mathrm{deg}$ with the `tight' prior. We see the posteriors approximately follow the priors for $R_0$, $\mu_{\ell,A*}$ and $\lambda$. The residuals in $\Omega_\mathrm{p}$ with respect to the model clearly have small-scale systematic variations and we see for $\ell<3.5\,\mathrm{deg}$ the pattern speed is biased high.

\begin{figure}
    \centering
    \includegraphics[width=\columnwidth]{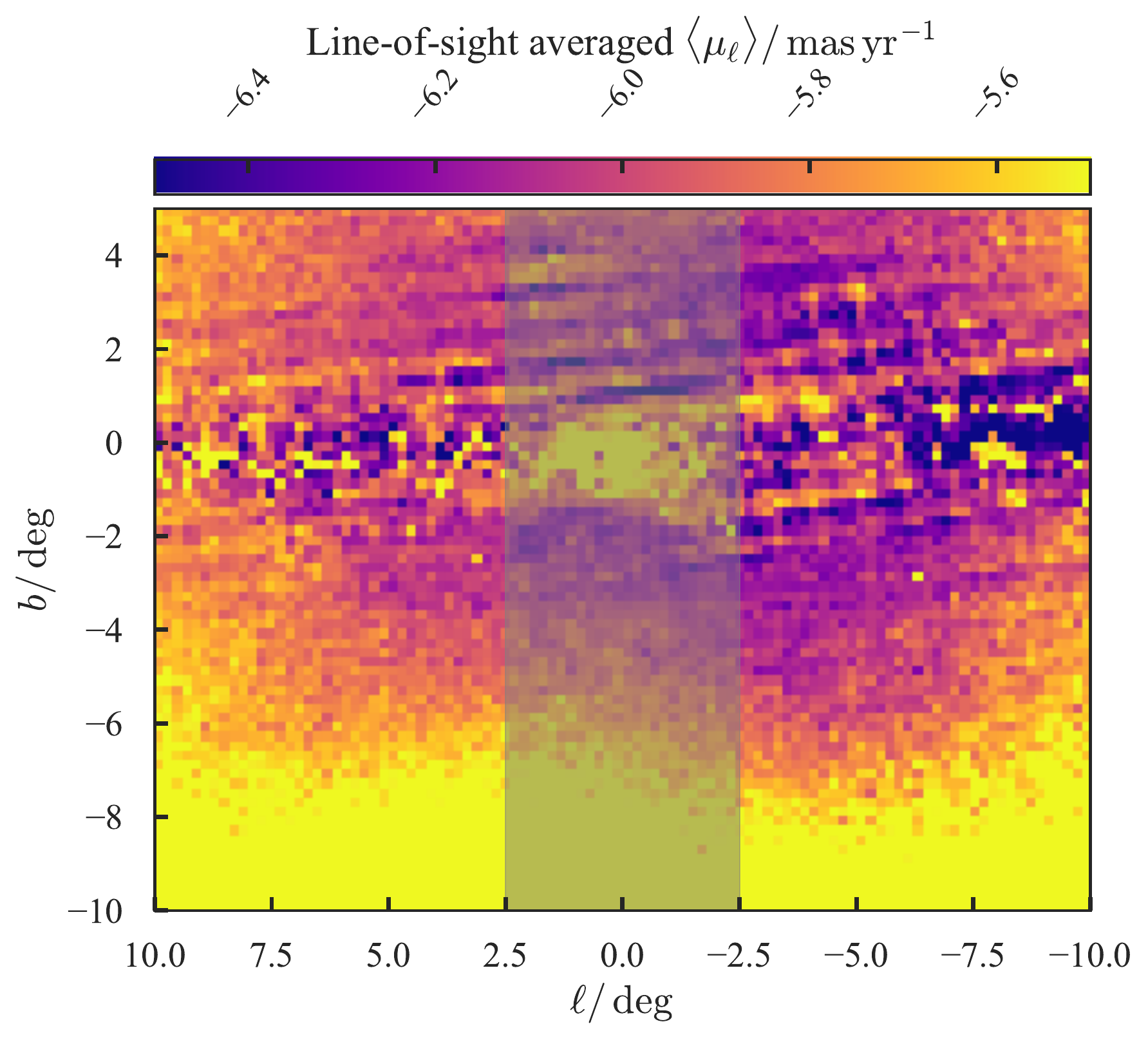}
    \includegraphics[width=\columnwidth]{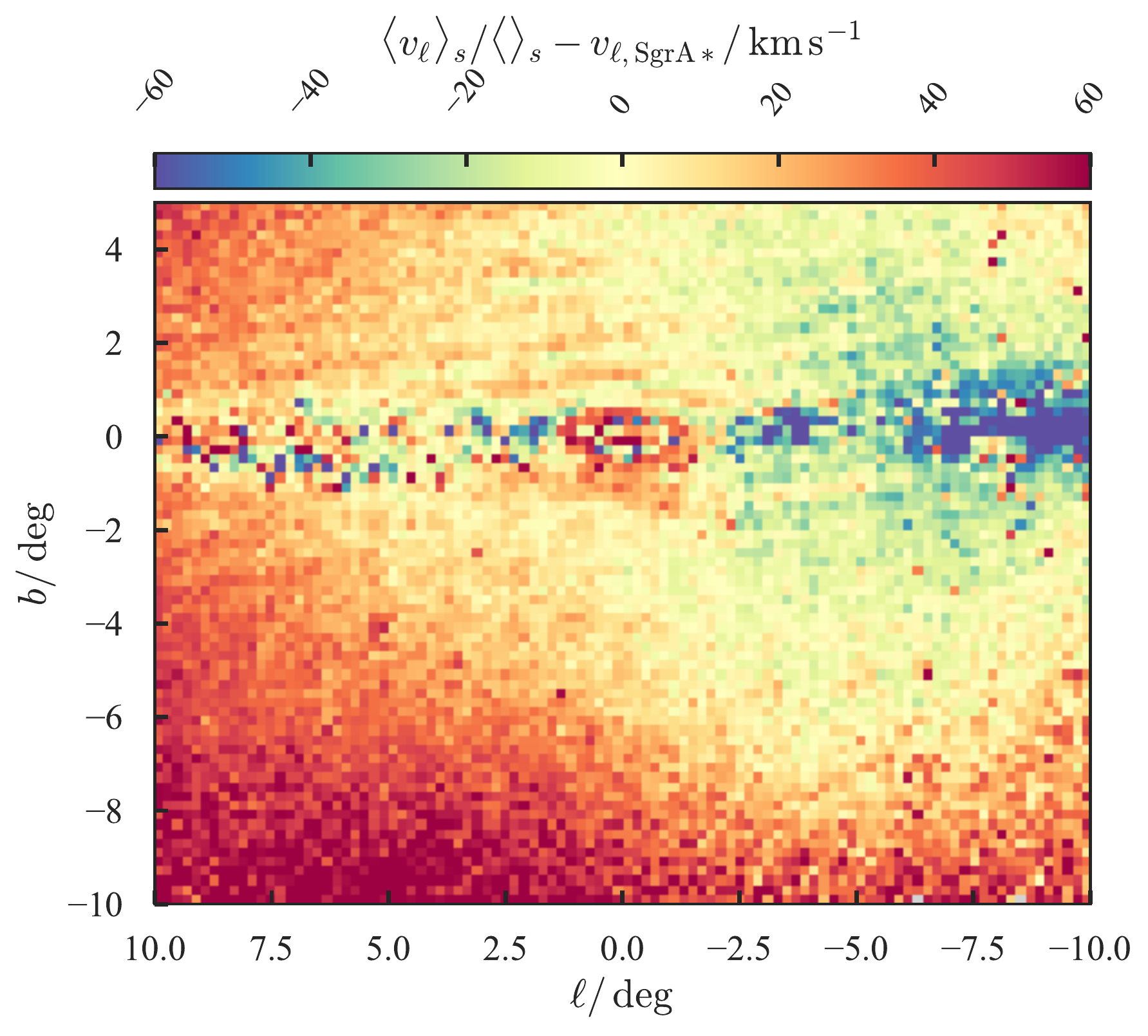}
    \caption{On-sky line-of-sight averaged mean $\mu_\ell$ (weighted by distance pdf) and mean $\ell$ velocity relative to Sgr A* (weighted by $s\rho$ as in the pattern speed estimator). Near the plane for $\ell<-2.5\,\mathrm{deg}$, the Gaia scanning law is clearly visible giving rise to proper motion systematics of $\sim0.5-1\,\mathrm{mas\,yr}^{-1}$. The grey shaded region shows $|\ell|<2.5\,\mathrm{deg}$ which we ignore in some of our fits.}
    \label{fig:meanl}
\end{figure}

It is puzzling why $\ell<0$ produces poor results. One reason could be that we do not have sufficient coverage in distance. It appears the pattern speed is increased if the nearby data (distance less than $6\,\mathrm{kpc}$) is cut out. This hints that there is insufficient background disc to counteract the foreground disc which is biasing the signal. We have experimented with using a fixed distance range of data relative to the bar major axis, but no one distance cut can reliably be chosen over any other. A further concern is unreliable extinction estimates which bias the velocities via incorrect distance estimates. We are using a 2d extinction map which is poor near the plane. For $\ell\ll0$ the bar gets closer to the plane due to geometric effects making this a bigger problem than for $\ell>0$. In the bottom section of Table~\ref{Table::results} we have removed $|b|<1\,\mathrm{deg}$ which produces very similar results -- a pattern speed of $\sim31\,\mathrm{km\,s}^{-1}\mathrm{kpc}^{-1}$ and a low $-\mu_{\ell,A*}$ -- although the pattern speed is slightly higher (see later simulation tests). 
Furthermore, in the analysis a proper motion systematic is degenerate with the proper motion of Sgr A*. In Fig.~\ref{fig:meanl} we show the mean proper motion averaged along the line-of-sight (accounting for the incompleteness effects). We see the pattern of the Gaia scanning law and associated systematic variations of the mean proper motion of up to $0.5-1\,\mathrm{mas\,yr}^{-1}$. These systematics are particularly bad for $\ell<-2.5\,\mathrm{deg}$ but appear less severe in the region $\ell>2.5\,\mathrm{deg}$ where our estimator is performing better. We have tested that this systematic pattern is present irrespective of using Gaia data in addition to VIRAC and irrespective of different cuts on proper motion quality. It is also independent of magnitude suggesting it arises from the relative-to-absolute correction for the VIRAC v1.1 (see Paper I).

In conclusion, our `best' model as measured by $\chi^2/N$ is for $\ell>2.5\,\mathrm{deg}$ with any choice of prior (all produce similar results). This yields a pattern speed of $\Omega_\mathrm{p}=(41\pm3)\mathrm{km\,s}^{-1}\mathrm{kpc}^{-1}$. Assuming a $v$ peculiar motion of the Sun of $12.24\,\mathrm{km\,s}^{-1}$ \citep{Schoenrich2010} and no non-axisymmetric streaming in the solar neighbourhood, this yields a corotation radius of 
$(5.7\pm0.4)\,\mathrm{kpc}$. 
We find an identical result if we instead use the rotation curve from \cite{Eilers2019}.

Our errorbars here are formal and likely underestimate the true uncertainty, particularly as systematic issues haven't enabled a consensus on the pattern speed to be formed across the observed volume and we know the uncertainties in the mean velocities are underestimated as we have not fully propagated the uncertainty in the density field or in the luminosity function. In Appendix~\ref{appendix:lf} we discuss our choice of luminosity function and demonstrate a broader red clump magnitude distribution ($0.12\,\mathrm{mag}$ instead of $0.067\,\mathrm{mag}$) gives near identical results.

\begin{table*}
    \caption{Results of application of our method to the data. The first column describes the subset of data used. Second column the priors employed (`Tight' is $R_0=(8.12\pm0.03)\,\mathrm{kpc}$ and $\mu_{\ell,A*}=(-6.379\pm0.026)\,\mathrm{mas\,yr}^{-1}$, `Loose' is $R_0=(8.2\pm0.09)\,\mathrm{kpc}$ and $\mu_{\ell,A*}=(-6.379\pm0.026)\,\mathrm{mas\,yr}^{-1}$, `No $R_0$' is $R_0=(8.12\pm2)\,\mathrm{kpc}$ and $\mu_{\ell,A*}=(-6.379\pm0.026)\,\mathrm{mas\,yr}^{-1}$ and `No $\mu_{\ell,A*}$' is $R_0=(8.12\pm0.03)\,\mathrm{kpc}$ and $\mu_{\ell,A*}=(-6.379\pm1)\,\mathrm{mas\,yr}^{-1}$. Subsequent columns give the median and standard deviation of the parameters. The final column gives the chi-squared per $\ell$ bin which is an approximate reflection of the goodness of fit.}
    \centering
    \begin{tabular}{c|c|c|c|c|c}
         Subset & Prior & $R_0/,\mathrm{kpc}$ & $-
         \mu_{\ell,A*}/\,\mathrm{mas\,yr}^{-1},$ & $\Omega_\mathrm{p}/\,\mathrm{km\,s}^{-1}\mathrm{kpc}^{-1}$ & $\chi^2/N$ \\
         \hline
All&Tight&$8.12\pm0.03$&$6.30\pm0.03$&$37.20\pm3.33$&$19.70$\\
&Loose&$8.21\pm0.08$&$6.31\pm0.03$&$36.85\pm3.09$&$12.32$\\
&No $R_0$&$8.24\pm0.14$&$6.30\pm0.03$&$37.07\pm3.21$&$11.36$\\
&No $\mu_{\ell,A*}$&$8.14\pm0.03$&$6.13\pm0.04$&$23.75\pm2.71$&$13.64$\\
\\
$\ell<0$&Tight&$8.12\pm0.03$&$6.29\pm0.03$&$27.08\pm1.96$&$19.58$\\
&Loose&$8.25\pm0.08$&$6.36\pm0.02$&$22.87\pm1.92$&$19.11$\\
&No $R_0$&$9.42\pm0.45$&$6.35\pm0.03$&$34.65\pm5.16$&$6.45$\\
&No $\mu_{\ell,A*}$&$8.13\pm0.03$&$6.31\pm0.05$&$26.47\pm3.03$&$20.83$\\
\\
$\ell>0$&Tight&$8.13\pm0.03$&$6.24\pm0.01$&$30.77\pm1.16$&$13.52$\\
&Loose&$8.22\pm0.06$&$6.24\pm0.01$&$30.81\pm1.20$&$8.93$\\
&No $R_0$&$8.24\pm0.08$&$6.24\pm0.01$&$30.88\pm1.18$&$7.31$\\
&No $\mu_{\ell,A*}$&$8.13\pm0.03$&$6.20\pm0.01$&$30.94\pm1.11$&$11.41$\\
\\
$\ell>2.5$&Tight&$8.12\pm0.03$&$6.36\pm0.02$&$42.09\pm2.50$&$1.42$\\
&Loose&$8.19\pm0.08$&$6.37\pm0.02$&$41.17\pm2.60$&$1.37$\\
&No $R_0$&$8.18\pm0.15$&$6.36\pm0.02$&$41.25\pm2.93$&$1.34$\\
&No $\mu_{\ell,A*}$&$8.12\pm0.03$&$6.32\pm0.04$&$39.41\pm3.63$&$1.51$\\
\\
$\ell<-2.5$&Tight&$8.13\pm0.03$&$6.37\pm0.02$&$23.26\pm1.75$&$3.44$\\
&Loose&$8.25\pm0.08$&$6.36\pm0.02$&$22.87\pm1.92$&$3.74$\\
&No $R_0$&$9.42\pm0.45$&$6.35\pm0.03$&$34.65\pm5.16$&$2.63$\\
&No $\mu_{\ell,A*}$&$8.13\pm0.03$&$6.31\pm0.05$&$26.47\pm3.03$&$2.60$\\
\\
$|\ell|>2.5$&Tight&$8.14\pm0.03$&$6.26\pm0.01$&$30.92\pm0.82$&$2.98$\\
&Loose&$8.24\pm0.06$&$6.26\pm0.01$&$30.96\pm0.78$&$2.70$\\
&No $R_0$&$8.28\pm0.08$&$6.26\pm0.01$&$31.00\pm0.77$&$2.58$\\
&No $\mu_{\ell,A*}$&$8.14\pm0.03$&$6.24\pm0.01$&$30.86\pm0.72$&$2.24$\\
\\
$|\ell|>2.5,|b|>1$&Tight&$8.13\pm0.03$&$6.21\pm0.01$&$31.70\pm0.69$&$4.13$\\
&Loose&$8.22\pm0.06$&$6.21\pm0.01$&$31.70\pm0.72$&$4.03$\\
&No $R_0$&$8.24\pm0.08$&$6.21\pm0.01$&$31.76\pm0.68$&$4.11$\\
&No $\mu_{\ell,A*}$&$8.13\pm0.03$&$6.19\pm0.01$&$31.29\pm0.64$&$3.81$\\

    \end{tabular}
    \label{Table::results}
\end{table*}

\section{Recovery of the pattern speed from a simulation}\label{Section::Recovery}
We provide a series of tests of our method for measuring the pattern speed from proper motion data using the continuity equation by application to a simulation. We first describe the simulation considered.

\subsection{Reference simulation}\label{Section::Simulation}

\begin{figure}
    \centering
    \includegraphics[width=\columnwidth]{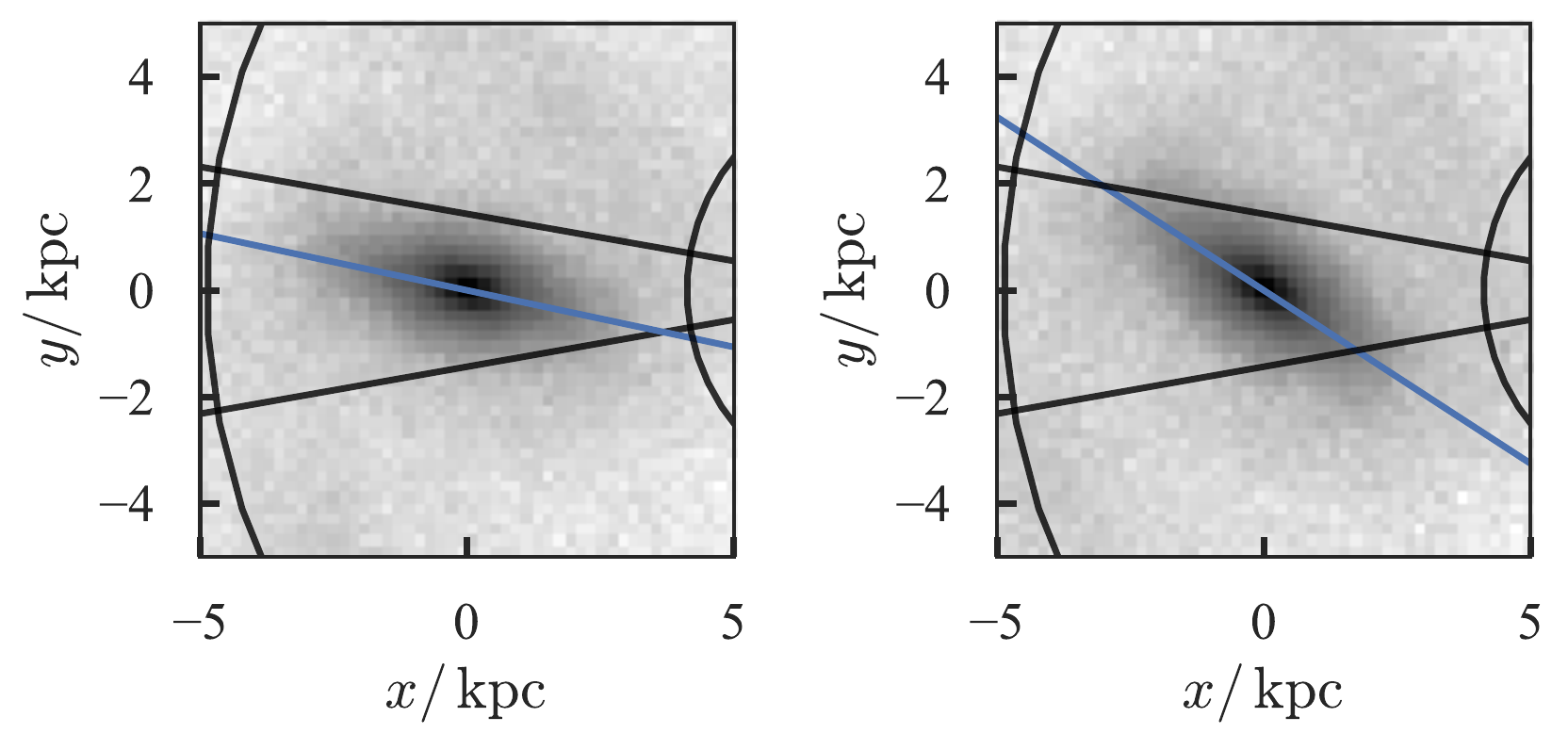}
    \caption{Top-down view of the reference simulation at two consecutive snapshots. The displayed particles have $|z|<2\,\mathrm{kpc}$. Black lines delineate the VVV footprint and the blue line shows the major axis of the bar. The two snapshots are $\sim8\,\mathrm{Myr}$ apart.}
    \label{fig:simulation_diagram}
\end{figure}

For the interpretation and testing of our results, we construct a simple reference simulation of a barred galaxy. This simulation is designed to approximately match the properties of the Milky Way although not to the level of detail of a full Made-to-Measure model \citep{Portail2017}. We use the initial condition generation \texttt{mkgalaxy} from \cite{McMillanDehnen2007}. The galaxy has three components: a disc, a bulge and a dark halo. We use the standard parameters: a \cite{Dehnen1999} disc with scale-length $R_d=1$, scale-height $z_d=0.1$ and mass $M_d=1$, a spherical \cite{Hernquist1990} bulge with scale-length $R_b=0.2$ and mass $M_b=0.2$, and a spherical \cite{NFW} halo with scale-length $R_h=6$ and mass $M_h=24$. The disc contains $200,000$ particles, the bulge $40,000$ and the halo $1,200,000$. The disc has a Toomre $Q$ parameter of $1.2$ making it radially-unstable to bar formation. Upon evolution for $200$ time units with \texttt{gyrfalcON} \citep{Dehnen2000}, the disc rapidly forms a bar that slows to $\Omega_\mathrm{p}\approx0.4$ by the end of the simulation. We measure the pattern speed from consecutive snapshots using the angular velocity of the second axis of the moment of inertia tensor for particles within $1.5$ simulation units of the centre. Using all particles and only those between $0.2$ and $0.5$ simulation units away from the midplane gives very similar results after $70$ simulation time units. We scale the final snapshot such that the scale-length of the disc is $2.5\,\mathrm{kpc}$ and the circular velocity of the disc $230\,\mathrm{km\,s}^{-1}$. This produces a bar rotating with pattern speed $\Omega_\mathrm{p}=45\,\mathrm{km\,s^{-1}\,kpc^{-1}}$ which we view at an angle of $\sim33\,\mathrm{deg}$ relative to the major axis. In Fig.~\ref{fig:simulation_diagram} we show two consecutive snapshots from the simulation scaled to the Milky Way. 

\subsection{Application}\label{Section::Simulation_Application}

We consider particles $-10<\ell/\,\mathrm{deg}<10$, $-10<b/\,\mathrm{deg}<5$, $-6<(s-8.12)/\,\mathrm{kpc}<6$ for a solar location of $R_0=8.12\,\mathrm{kpc}$ viewing the bar at $\sim33\,\mathrm{deg}$. Proper motions are computed using the total solar velocity of $(u,v,w)_\odot=(11.1,-4.74\mu_{\ell,A*}R_0,7.25)\,\mathrm{km\,s}^{-1}$ and $\mu_{\ell,A*}=-6.379\,\mathrm{mas\,yr}^{-1}$.

When testing with simulations, we require expressions equivalent to those in Section~\ref{Section::PatternSpeed} but appropriate for a finite sampling of the underlying smooth functions. For a set of tracer particles, $\rho=\sum_im_i\delta^{(3)}(\bs{x}-\bs{x}_i)$ we construct bins (indexed by $n$) in $\ell$, $b$ or $s$ depending on the free variable in the pattern speed expression. The bins are centred on e.g. $\ell_n$ with width $\Delta\ell$. We evaluate
\begin{equation}
    \Omega_\mathrm{p}(\ell_n) = \frac{\sum_i\,m_i v_{\ell,i}/(s_i\cos b_i)}{\sum_i
    m_i(R_0\cos\ell_i/(s_i\cos b_i)-1)},
\end{equation}
\begin{equation}
    \Omega_\mathrm{p}(b_n) = -\frac{\sum_i\,m_i v_{b,i}/s_i}{\sum_i
   m_i R_0 \sin\ell_i\sin b_i/s_i},
\end{equation}
\begin{equation}
    \Omega_\mathrm{p}(s_n) = \frac{\sum_i\,m_i v_{||,i}/s_i}{\sum_i
    m_iR_0 \sin\ell_i\cos b_i/s_i},
\end{equation}
where velocities are in the Galactic rest frame.
As in Section~\ref{Section::PatternSpeed}, we write the first of these expressions as
\begin{equation}
\begin{split}
       \Omega_\mathrm{p}(\ell_n) &= \frac{v_\odot-u_\odot\tan\ell_n+\mathcal{F}}{R_0-\mathcal{K}},\\\mathcal{F}&= \frac{\sum_i\,m_iv'_{\ell,i}/(s_i\cos b_i)}{\sum_i\,m_i\cos\ell_i/(s_i\cos b_i)},
       \>\mathcal{K}= \frac{\sum_i\,m_i}{m_i\cos\ell_i/(s_i\cos b_i)},
\end{split}
\end{equation}

We use the probabilistic model in equation~\eqref{Eqn::OmegaModel} to infer $|\mu_{\ell,A*}|$, $R_0$, $\Omega_\mathrm{p}$ and the distance systematic factor $\lambda$. The data covariance matrix $\bs{\Sigma}_{\mathcal{X}i}$ is computed using $100$ bootstrap resamples of the particle properties in each bin.

\begin{figure}
    \centering
    \includegraphics[width=\columnwidth]{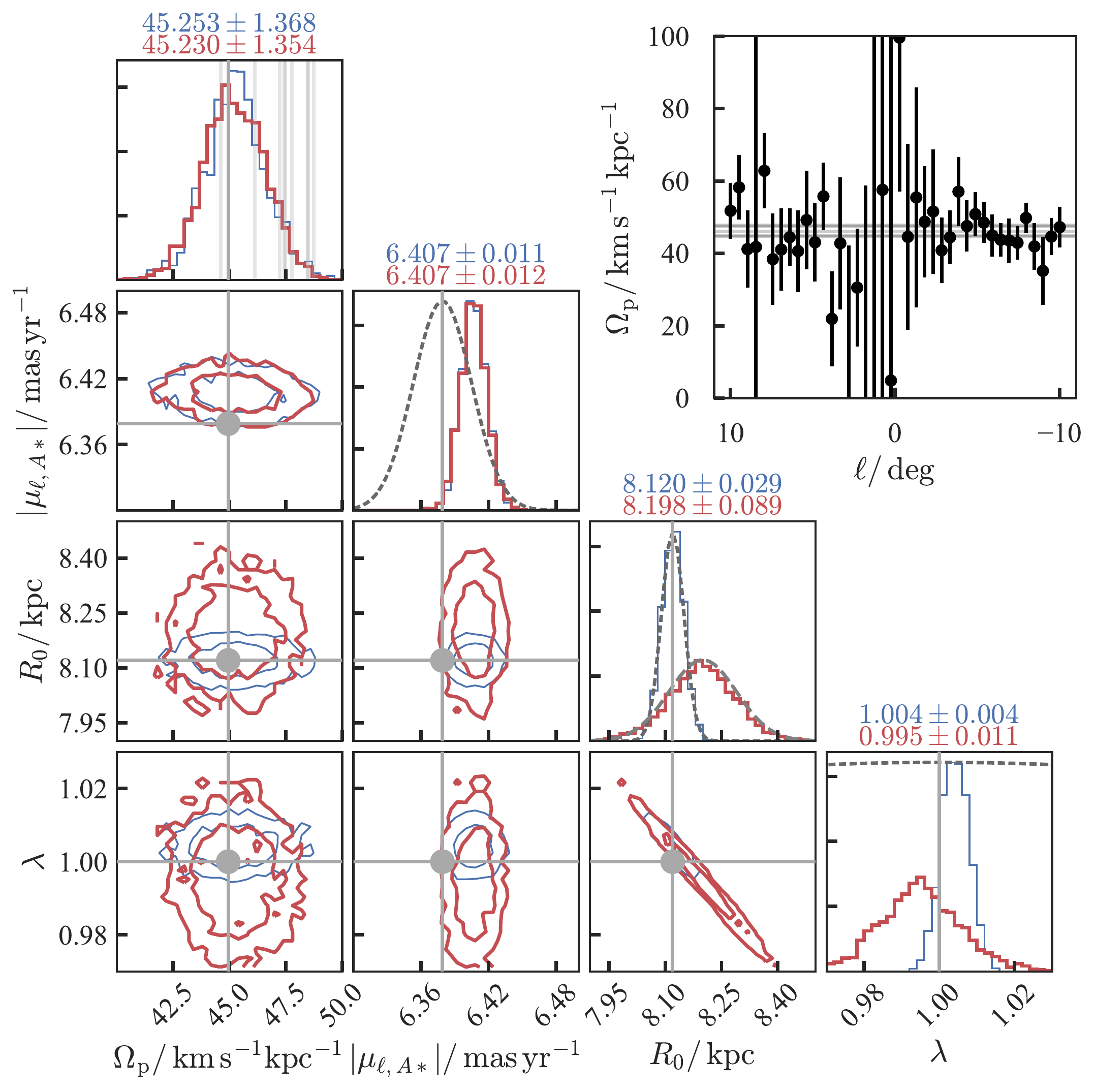}
    \caption{Inference of the pattern speed from simulation: corner plot shows the recovery of the pattern speed using two different priors on $R_0$ (blue from \protect\cite{McMillan2017} and red from \protect\cite{GravityCollab2018}). Above each panel we show the mean and standard deviation inferred for each parameter. The vertical/horizontal lines show the true values (in the top left panel we show the estimate of $\Omega_\mathrm{p}$ from the last $10$ snapshots). The inset shows the pattern speed estimator in bins of Galactic longitude for the true parameters. Horizontal lines correspond to those in the top left corner panel. $\lambda$ is a fractional distance systematic.}
    \label{fig:simulation_recovery}
\end{figure}

In Fig.~\ref{fig:simulation_recovery}, we show the inference for our model using the two Galactic centre distance priors. We have not applied any distance systematic to the simulation data, we use bins in $\ell$ of width $\Delta\ell=0.5\,\mathrm{deg}$ and we set $N_\mathrm{max}=2$. The inset shows the $\Omega_\mathrm{p}(\ell)$ estimate for each bin in $\ell$. We note that the central regions produce noisy estimates of $\Omega_\mathrm{p}$. This is probably because both numerator and denominator in the estimator are small, but could also be because the considered stars form part of the original bulge component which is perhaps not rotating with the bar. We also see that negative Galactic longitude produces more precise $\Omega_\mathrm{p}$ estimates as there are more stars in the solid angle considered. From the inference, we find that with both priors the pattern speed is well recovered (the expected pattern speed is $45\,\mathrm{km\,s}^{-1}\mathrm{kpc}^{-1}$). The solar radius posterior follows the adopted prior, the proper motion of Sgr A* is tighter than the prior and the distance systematic is recovered as unbiased.

\begin{figure}
    \centering
    \includegraphics[width=\columnwidth]{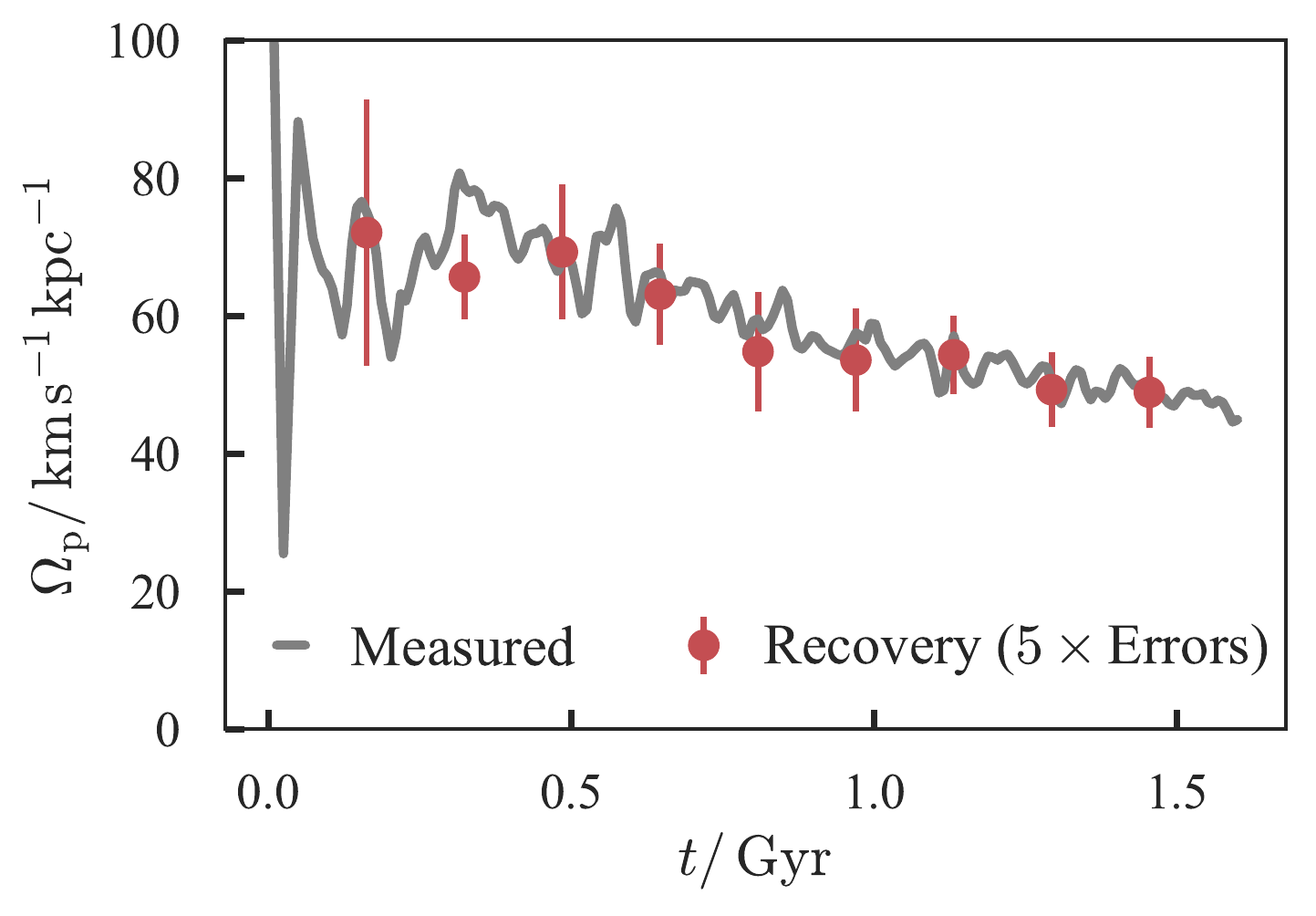}
    \caption{Recovery of $\Omega_\mathrm{p}$ from our reference simulation at different times. The grey line shows the pattern speed inferred from the moments of inertia of consecutive snapshots. Red errorbars show the recovery from our modelling where we have multiplied the uncertainties by $5$ for visibility.}
    \label{fig:simulation_recovery_times}
\end{figure}

We apply our method to $9$ snapshots from the simulation where for each snapshot the bar is rotated to an angle $33\,\mathrm{deg}$ with respect to the line-of-sight. We then observe simulation particles within the VVV bulge region and use $N_\mathrm{max}=2$, $\Delta\ell=0.5\,\mathrm{deg}$ and $(\mu_{R0},\sigma_{R0})=(8.2,0.09)\,\mathrm{kpc}$. The results are shown in Fig.~\ref{fig:simulation_recovery_times}. We observe the decaying pattern speed of the bar. The recovery is shown with red errorbars (multiplied by $5$ for visibility). At all snapshots we recover the pattern speed with increasing precision at later times when the bar is more established. At early times, transient phenomena cause more uncertainty in the pattern speed. For $t<0.5\,\mathrm{Gyr}$ the pattern speed measured for all particles and those between $0.2$ and $0.5$ simulation units of the plane disagree slightly suggesting the bar hasn't reached equilibrium yet.

\begin{figure}
    \centering
    \includegraphics[width=\columnwidth]{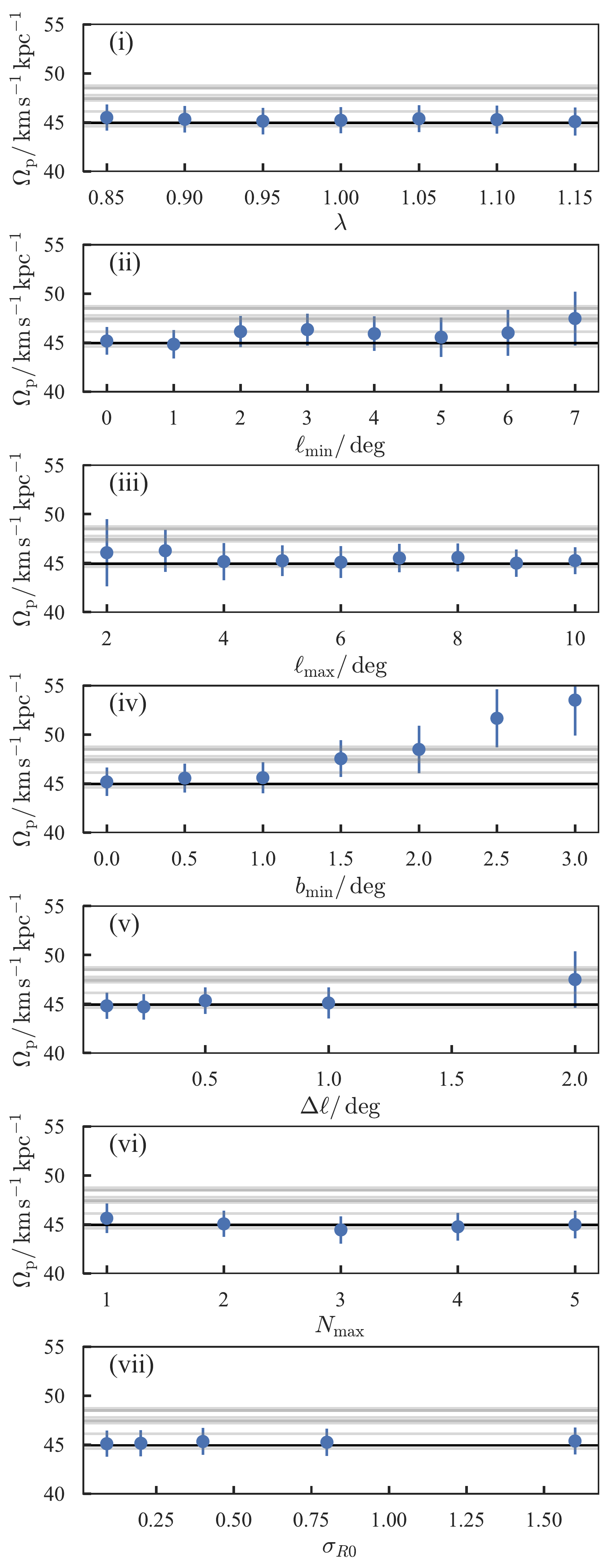}
    \caption{Recovery of $\Omega_\mathrm{p}$ from our reference simulation using different setups: (i) including a relative distance systematic, (ii) using data with $|\ell|>\ell_\mathrm{min}$, (iii) using data with $|\ell|<\ell_\mathrm{max}$, (iv) using data with $|b|>b_\mathrm{min}$, (v) varying the bin width $\Delta\ell$, (vi) varying the number of polynomial coefficients in the model and (vii) increasing the width of the prior on $R_0$. The dark solid line is the `true' pattern speed and the fainter lines the pattern speed from the last $10$ snapshots.}
    \label{fig:simulation_recovery_parameters}
\end{figure}

We perform further experiments varying (i) the systematic distance bias used to construct the mock data, (ii) the minimum $|\ell|$ considered, (iii) the maximum $|\ell|$ considered, (iv) the minimum $|b|$ considered, (v) the bin widths, (vi) number of polynomial terms in the model $N_\mathrm{max}$ and (vii) the width of the solar radius prior. The results are shown in Fig.~\ref{fig:simulation_recovery_parameters}. It is satisfying that varying most parameters does not bias the pattern speed significantly. Both distance systematics and uncertainty in the distance to the Galactic centre are irrelevant to the recovery (due to the degeneracy compensating one for the other). Using data with $|\ell|>\ell_\mathrm{min}$ or $|\ell|<\ell_\mathrm{max}$ does not alter the results other than increasing the uncertainty when less data is used (similar result for the bin size $\Delta\ell$) -- this is expected as the estimator $\Omega_\mathrm{p}(\ell)$ does not require coverage in $\ell$. The same is not true when considering only data with $|b|>b_\mathrm{max}$. The pattern speed is systematically biased when in-plane data is excluded. We previously checked the pattern speed of stars between $0.2$ and $0.5$ simulation units of the plane (approximately $b>3\,\mathrm{deg}$) was near identical to using all particles. It appears the recovery is satisfactory for $b_\mathrm{max}\lesssim1.5\,\mathrm{deg}$.

We have also attempted to use the $\Omega_\mathrm{p}(b)$ estimator on the simulation but we find it is not reliable. This is possibly due to lack of resolution but also could be due to the boundary terms dominating the signal.

To summarise, we find that the recovery of $\Omega_\mathrm{p}$ is not affected by distance systematics, bin sizes, number of polynomial terms used $N_\mathrm{max}$, the range of $\ell$ used and the prior on $R_0$. If we filter out low latitude data, $|b|<b_\mathrm{min}$, we find the results are biased if $b_\mathrm{min}\gtrsim1.5\,\mathrm{deg}$. $R_0$ is completely degenerate with a distance systematic.

\subsection{Boundary terms}

When deriving the estimators for the pattern speed, we removed terms by arguing that they vanish at the boundaries. In realistic applications we are unable to integrate over all space so our estimators are biased by the non-zero contributions of the boundary terms. A boundary term near constant in $\ell$ will produce a constant offset in the Galactic centre distance or the motion of Sgr A*, whereas the unlikely case where boundary contributions behave like uncorrelated noise in $\ell$ will not bias the results due to our modelling excess scatter in $\mathcal{F}$ and $\mathcal{K}$. The concerning case is for near-linear variation in the boundary terms which masquerade as a change in the pattern speed.

When deriving the $\ell$-estimator, there are six boundary terms we discount:
\begin{equation}
    \begin{split}
       &(1)\>\>\>\>\>\>\>\>\Big[\int_{-\pi}^{\ell}\mathrm{d}\ell\int_{b_\mathrm{min}}^{b_\mathrm{max}}\mathrm{d}b\,s^2\rho v_{||}\cos b\Big]_{s_\mathrm{min}}^{s_\mathrm{max}},\\
       &(2)\>\>\>\>\>\>\>\>\Big[\int_{-\pi}^{\ell}\mathrm{d}\ell\int_{s_\mathrm{min}}^{s_\mathrm{max}}\mathrm{d}s\,s\rho v_{b}\cos b\Big]_{b_\mathrm{min}}^{b_\mathrm{max}},\\
       &(3)\>\>\>-\Big[\int_{-\pi}^{\ell}\mathrm{d}\ell\int_{b_\mathrm{min}}^{b_\mathrm{max}}\mathrm{d}b\,s^2\rho \cos^2b\sin\ell\Big]_{s_\mathrm{min}}^{s_\mathrm{max}}\Omega_\mathrm{p}R_0,\\
       &(4)\>\>\>\>\>\>\>\>\Big[\int_{-\pi}^{\ell}\mathrm{d}\ell\int_{s_\mathrm{min}}^{s_\mathrm{max}}\mathrm{d}s\,s\rho \sin b\cos b\sin\ell\Big]_{b_\mathrm{min}}^{b_\mathrm{max}}\Omega_\mathrm{p}R_0,\\
       &(5)\>\>\>-\Big[\int_{b_\mathrm{min}}^{b_\mathrm{max}}\int_{s_\mathrm{min}}^{s_\mathrm{max}}\mathrm{d}s\,s\rho(R_0\cos\ell-s\cos b)\Big]_{\ell=-\pi}\Omega_\mathrm{p},\\
       &(6)\>\>\>-\Big[\int_{b_\mathrm{min}}^{b_\mathrm{max}}\int_{s_\mathrm{min}}^{s_\mathrm{max}}\mathrm{d}s\,s\rho v_\ell\Big]_{\ell=-\pi}.
    \end{split}
\end{equation}
The first of these terms involves the line-of-sight velocities so in the absence of full spectroscopic coverage of the sky we must use simulations to estimate its amplitude. The second term involves the latitudinal velocities so in theory could be estimated from proper motion data, except we require proper motion data outside the observational volume to evaluate the $\ell$ integral. 
The third and fourth terms involve no velocities so can be evaluated from the data modulo the same considerations about integrating over $\ell$. The final two terms are the lower limits of the $\ell$ integrals which cannot be evaluated from the data. 

\begin{figure}
    \centering
    \includegraphics[width=\columnwidth]{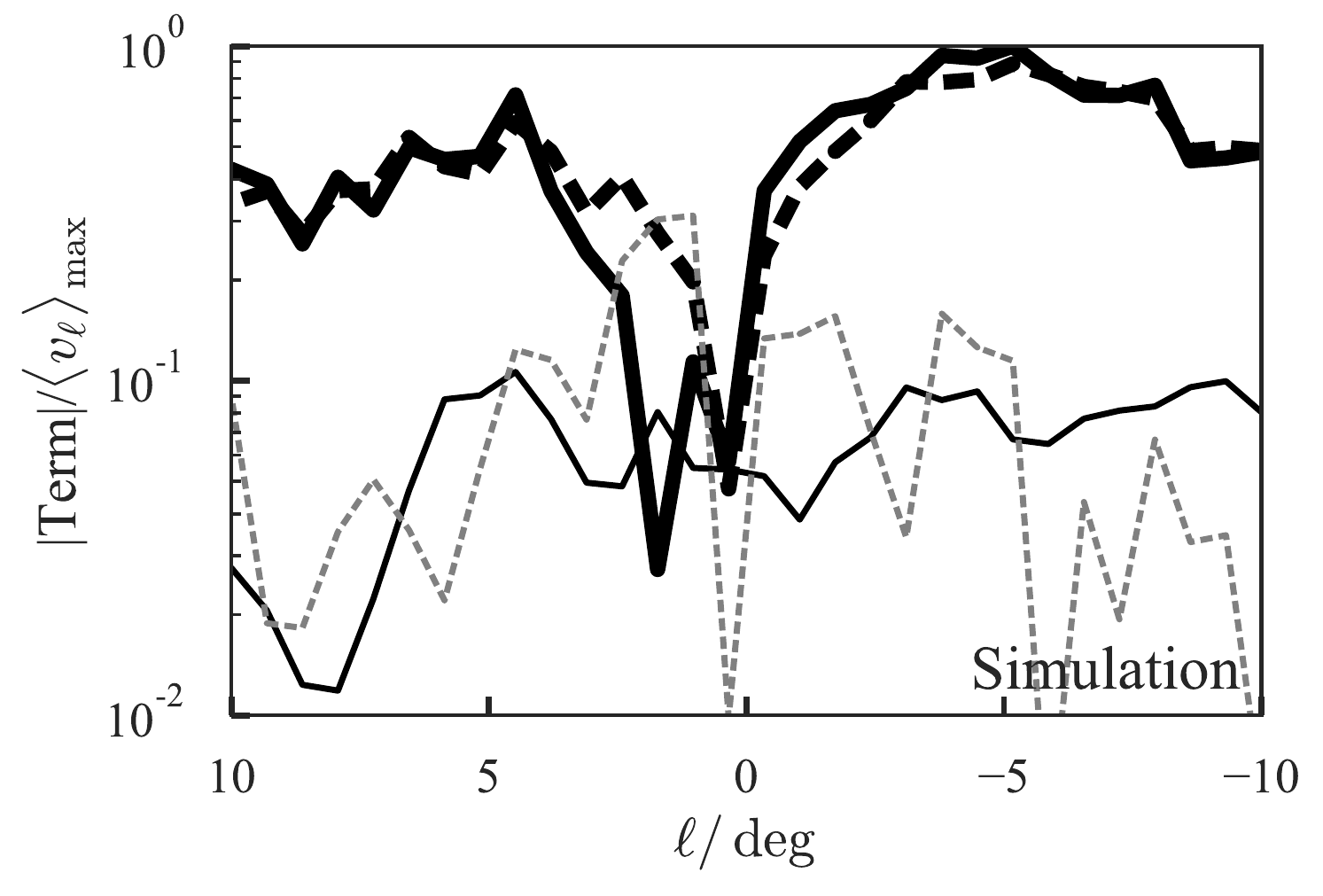}
    \caption{Amplitude of the boundary terms neglected in our derivation of the pattern speed estimator $\Omega_\mathrm{p}(\ell)$. The thick black lines show the two terms used in the estimator ($\langle v_\ell \rangle_{b,s}$ [solid] and $\Omega_\mathrm{p}{\langle R_0\cos\ell-s\cos b\rangle_{b,s}}$ [dashed]), the grey line is their difference and the thinner black line is the sum of the neglected boundary terms.
    The neglected boundary terms are of order $10\percent$ the terms used in the estimator.}
    \label{fig:boundary_terms}
\end{figure}

As we are unable to truly estimate the boundary terms from the data, we instead use our reference simulation. We convert the integrals into sums over particles as, for example,
\begin{equation}
\begin{split}
    &\Big[\int_{-\pi}^{\ell}\mathrm{d}\ell\int_{b_\mathrm{min}}^{b_\mathrm{max}}\mathrm{d}b\,s^2\rho v_{||}\cos b\Big]_{s_\mathrm{min}}^{s_\mathrm{max}}\\&\approx\frac{1}{2\Delta s}\Big[\int_{s-\Delta s}^{s+\Delta s}\mathrm{d}s'\,\int_{-\pi}^{\ell}\mathrm{d}\ell\int_{b_\mathrm{min}}^{b_\mathrm{max}}\mathrm{d}b\,s'^2\rho v_{||}\cos b\Big]_{s=s_\mathrm{min}}^{s=s_\mathrm{max}}\\&=\frac{1}{2\Delta s}\Big[\sum_i m_i\int_{s-\Delta s}^{s+\Delta s}\mathrm{d}s'\,\int_{-\pi}^{\ell}\mathrm{d}\ell\int_{b_\mathrm{min}}^{b_\mathrm{max}}\mathrm{d}b s'^2 v_{||}\cos b\delta(\bs{x}\!-\!\bs{x}_i)\Big]_{s=s_\mathrm{min}}^{s=s_\mathrm{max}}\\&=\frac{1}{2\Delta s}\Big[\sum_{i,(\ell_i,b_i,s_i)\in\mathcal{V}} m_i v_{i,||}\Big]_{s=s_\mathrm{min}}^{s=s_\mathrm{max}},
\end{split}
\end{equation}
where the sum is over the particles inside the volume $\mathcal{V}$ defined by $\ell=(-\pi,\ell), b=(b_\mathrm{min},b_\mathrm{max}),s'=(s-\Delta s,s+\Delta s)$. We set $\Delta s=1\,\mathrm{kpc}$ and $\Delta b=0.2\,\mathrm{deg}$.

In Fig.~\ref{fig:boundary_terms}, we show the amplitude of the boundary terms estimated from the reference simulation. 
We use the pattern speed measured from consecutive snapshots. 
We observe
the sum of the boundary terms is of order the difference in the estimator quantities and is approximately $10\percent$ the magnitude of the estimator quantities. We see that near $\ell=0$ the boundary terms are significant relative to the estimator quantities as $v_\ell$ approaches zero here. This corresponds to the poor estimates of $\Omega_\mathrm{p}$ seen in Fig.~\ref{fig:simulation_recovery}. We have found that individual terms (1)-(4) in the sum of the boundary terms can be of order the estimator quantities but their sum is much smaller. When deriving the estimator formulae we assumed each of the terms was small but this does not appear to be true. It is perhaps fortuitous that their sum is negligible, but this appears to explain the degree of accuracy obtained through application of estimators to the simulation. The $b$-boundary terms can be made smaller if a symmetric interval is used, as we have done in the analysis of the data by assuming symmetry in $b=0$. Employing $b_\mathrm{max}=10\,\mathrm{deg}$ instead of $5\,\mathrm{deg}$ reduces the sum of the boundary terms to $\lesssim1\percent$ for $\ell<0$ and $2-8\percent$ for $\ell>0$. Additionally, we observe that the individual boundary terms, as well as their sum, are near constant with $\ell$ so will lead to systematic offsets in the properties of the Galactic centre rather than the pattern speed.

\section{Conclusions}\label{Section::Conclusions}

We have measured the pattern speed of the Milky Way bar as $\Omega_\mathrm{p}=(41\pm 3)\,\mathrm{km\,s^{-1}\,kpc^{-1}}$ using proper motion data from VVV and Gaia DR2.  This places corotation at 
$(5.7\pm0.4)\,\mathrm{kpc}$. This result was obtained from the more reliable near-side of the bar and when the entire bar region is considered we obtain $\Omega_\mathrm{p}=(31\pm 1)\,\mathrm{km\,s^{-1}\,kpc^{-1}}$ but an inconsistent position and velocity of the Galactic centre. This suggests systematic uncertainties in our measurement of  $5-10\,\mathrm{km\,s^{-1}kpc^{-1}}$.

To establish this, we developed new estimators for the pattern speed using transverse velocity data derived from the Galactic proper motion components. These estimators use the Tremaine-Weinberg method of integrating the continuity equation. Our new estimators are tailored for use specifically in the Milky Way. Using our longitudinal velocity estimator, we build a probabilistic model that allows for full propagation of uncertainties. We have demonstrated the performance of the method through application to a disc galaxy simulation that has formed a dynamical bar. Although we only consider a selection of the simulation comparable to the VVV survey volume, we find our method robustly recovers the pattern speed at a number of simulation times. The only biases we detect are when excluding in-plane stars $|b|\lesssim1.5\,\mathrm{deg}$ when the method overestimates the pattern speed. When applying to data, only fields with $\ell>2.5\,\mathrm{deg}$ appear to produce reliable estimates possibly due to insufficient distance coverage, extinction effects or proper motion systematics.

\section*{Acknowledgements}
JLS thanks the Science and Technology Facilities Council, the Leverhulme Trust, the Newton Trust and Christ's College, Cambridge for financial support. We acknowledge useful conversations with HongSheng Zhao and Ortwin Gerhard, and we thank the anonymous referee for a close reading of the paper. This research was supported in part by the National Science Foundation under Grant No. NSF PHY-1748958.

This paper made used of the Whole Sky Database (wsdb) created by Sergey Koposov and maintained at the Institute of Astronomy, Cambridge by Sergey Koposov, Vasily Belokurov and Wyn Evans with financial support from the Science \& Technology Facilities Council (STFC) and the European Research Council (ERC).

Based on data products from observations made with ESO Telescopes at the La Silla or Paranal Observatories under ESO programme ID 179.B-2002. This work has made use of data from the European Space Agency (ESA) mission
{\it Gaia} (\url{https://www.cosmos.esa.int/gaia}), processed by the {\it Gaia}
Data Processing and Analysis Consortium (DPAC,
\url{https://www.cosmos.esa.int/web/gaia/dpac/consortium}). Funding for the DPAC
has been provided by national institutions, in particular the institutions
participating in the {\it Gaia} Multilateral Agreement. This publication makes use of data products from the Two Micron All Sky Survey, which is a joint project of the University of Massachusetts and the Infrared Processing and Analysis Center/California Institute of Technology, funded by the National Aeronautics and Space Administration and the National Science Foundation. Funding for the Sloan Digital Sky Survey IV has been provided by the Alfred P. Sloan Foundation, the U.S. Department of Energy Office of Science, and the Participating Institutions. SDSS-IV acknowledges support and resources from the Center for High-Performance Computing at the University of Utah. The SDSS web site is www.sdss.org. SDSS-IV is managed by the Astrophysical Research Consortium for the  Participating Institutions of the SDSS Collaboration including the 
Brazilian Participation Group, the Carnegie Institution for Science, 
Carnegie Mellon University, the Chilean Participation Group, the French Participation Group, Harvard-Smithsonian Center for Astrophysics, 
Instituto de Astrof\'isica de Canarias, The Johns Hopkins University, 
Kavli Institute for the Physics and Mathematics of the Universe (IPMU) / 
University of Tokyo, the Korean Participation Group, Lawrence Berkeley National Laboratory, 
Leibniz Institut f\"ur Astrophysik Potsdam (AIP),  
Max-Planck-Institut f\"ur Astronomie (MPIA Heidelberg), 
Max-Planck-Institut f\"ur Astrophysik (MPA Garching), 
Max-Planck-Institut f\"ur Extraterrestrische Physik (MPE), 
National Astronomical Observatories of China, New Mexico State University, 
New York University, University of Notre Dame, 
Observat\'ario Nacional / MCTI, The Ohio State University, 
Pennsylvania State University, Shanghai Astronomical Observatory, 
United Kingdom Participation Group,
Universidad Nacional Aut\'onoma de M\'exico, University of Arizona, 
University of Colorado Boulder, University of Oxford, University of Portsmouth, 
University of Utah, University of Virginia, University of Washington, University of Wisconsin, 
Vanderbilt University, and Yale University.

This publication made use of the Python science stack: 
\texttt{numpy} \citep{numpy}, 
\texttt{scipy} \citep{scipy},
\texttt{matplotlib} \citep{matplotlib},
\texttt{ipython} \citep{ipython} and
\texttt{pandas} \citep{pandas}.




\bibliographystyle{mnras}
\bibliography{bibliography} 



\appendix

\section{Estimators accounting for the true Galactic plane}\label{Appendix:z0}
In Section~\ref{Section::PatternSpeed} we presented estimators for the pattern speed using Galactic coordinates assuming $b=0$ lies in the Galactic plane. The Sun's measured height above the disc plane of $z_0=(25\pm5)\,\mathrm{pc}$ \citep{BlandHawthornGerhard2016} means this approximation is probably sufficient. However, if one were to apply the expressions to the long thin bar \citep{Wegg2015} it may be necessary to incorporate this effect. Therefore, for completeness we present estimators accounting for the additional offset. Sgr A* lies at latitude $b_{A*}=-0.046\,\mathrm{deg}$ so taking $R_0=8.12\,\mathrm{kpc}$, the angle between midplane and $b=0$ is $\gamma\sim0.13\,\mathrm{deg}$ \citep[see Figure 5 of][]{BlandHawthornGerhard2016}. 

We return to equation~\eqref{eqn::continuity} in which the second term ($\nabla\cdot(\rho\bs{v})$ is invariant under rotations so unaffected by $z_0$). Keeping $(x,y,z)$ as the Cartesian coordinates aligned with $b=0$ we express the first bracket as
\begin{equation}
\begin{split}
    &\cos\gamma\Big(y\frac{\partial\rho}{\partial x}-x\frac{\partial\rho}{\partial y}\Big)-
    \sin\gamma\Big(y\frac{\partial\rho}{\partial z}-z\frac{\partial\rho}{\partial y}\Big)=\\ &\cos\gamma\Big(y\frac{\partial\rho}{\partial x}-x\frac{\partial\rho}{\partial y}\Big)-
    \sin\gamma\Big(-\frac{\cos\ell\sin b}{\cos b}\frac{\partial\rho}{\partial \ell}+\sin\ell\frac{\partial\rho}{\partial b}\Big).
\end{split}
\label{eqn::extraterms}
\end{equation}
The first term gives $\cos\gamma$ multipled by the terms in the regular estimator. As in Section~\ref{Section::PatternSpeed} we multiply by $s^2\cos b$ and we can rearrange the terms in the second bracket of equation~\eqref{eqn::extraterms} as
\begin{equation}
    -s^2\sin b\frac{\partial(\rho\cos\ell)}{\partial \ell}
    +s^2\sin \ell\frac{\partial(\rho\cos b)}{\partial b}.
\end{equation}
These additional terms are whole derivatives so we can proceed in the normal way deriving

\begin{equation}
\begin{split}
    \Omega_\mathrm{p}(\ell) &= \frac{\langle v_\ell \rangle_{b,s}}{\langle (R_0\cos\ell-s\cos b)\cos\gamma-s\cos\ell\sin b\sin\gamma\rangle_{b,s}},\\
    \Omega_\mathrm{p}(b) &= -\frac{\langle v_b\rangle_{\ell,s}}{
    \langle R_0\sin b\sin \ell\cos\gamma-s\sin\ell\cos b\sin\gamma\rangle_{\ell,s}},\\
    \Omega_\mathrm{p}(s) &= \frac{\langle v_{||}\cos b\rangle_{\ell,b}}{
    R_0\cos\gamma\langle \sin\ell\cos^2 b\rangle_{\ell,b}},
\end{split}
\label{eqn::gammaequations}
\end{equation}
where $R_0$ is the distance to Sgr A* (formally it is the distance to the intercept between the axis normal to the disc plane and $b=0$ which is $R_0\cos b_{A*}\approx R_0$). $1-\cos\gamma\approx5\times10^{-6}$ so it is sufficient to take $\cos\gamma=1$. Furthermore, the final term in the denominator of $\Omega_\mathrm{p}(\ell)$ will be approximately zero for a density distribution near symmetric in $b$. Therefore, as expected, the effect of non-zero $z_0$ on the estimators is small and will only produce a noticeable effect using the $b$-estimator if $b\sim\gamma$. 
A small further consideration is that Sgr A* is located at $\ell_{A*}=-0.056\,\mathrm{deg}$ which can be approximately accounted for by $\ell\rightarrow\ell-\ell_{A*}$ in the above expressions,  slightly shifting the centre of rotation. An alternative approach is using equation~\eqref{eqn::gammaequations} with `Galactic coordinates' centred on $(\ell,b)_{A*}$ and a rotation of $\gamma\approx z_0/R_0=0.176\,\mathrm{deg}$.

\section{Red giant luminosity function}\label{appendix:lf}

The modelling in Paper I rested on an appropriate model for the red giant luminosity function for the bulge stars. The luminosity function is necessary for measuring both the density structure and for converting proper motions into transverse velocity distributions. Due to our uncertainty in the luminosity function, there is a systematic uncertainty in the results presented in the main body of this paper. In Paper I we employed the luminosity function from \cite{Simion2017} computed using PARSEC isochrones \citep{PARSEC} and assuming a single age of $10\,\mathrm{Gyr}$ and a Gaussian in metallicity centred on $0\,\mathrm{dex}$ with a width of $0.4\,\mathrm{dex}$. Here we will briefly explore whether this luminosity function is appropriate and discuss how our results change when varying the luminosity function.

We test the validity of our luminosity function using stars in the solar neighbourhood and stars in bulge globular clusters. First, we take all stars in Gaia DR2 \citep{Gaia2016,Gaia2018} cross-matched with 2MASS \citep{Skrutskie2006} with \citep{SFD} $\mathrm{E}(B-V)<0.1$, \texttt{parallax\_over\_error}$>50$, \texttt{parallax}$>1$, $(J-K_s)_0>0.4$, $G>4$ and high quality 2MASS photometry (\texttt{ph\_qual}=`AAA' and \texttt{cc\_flg}=`000'). We de-redden the magnitudes using \citep{SFD} $\mathrm{E}(B-V)$ with coefficients from \cite{Yuan2013}. We match this catalogue to the `pristine' red clump stars from the catalogue of \cite{Ting2018}. We inspected the colour-magnitude diagrams of stars within $1.5r_\mathrm{h}$ of the centres of bulge globular clusters from \cite[][2010 version]{Harris1996} using a PSF version of the VIRAC catalogue (Smith et al., in prep.). We measured the proper motion as the peak of the 2d proper motion distribution and selected only stars within $2\,\mathrm{mas\,yr}^{-1}$ in $\mu_\alpha$ and $\mu_\delta$ of the peak. We found NGC 6553 had the clearest giant branch and also has a metallicity of $-0.18\,\mathrm{dex}$ making it an appropriate reference case. 

\begin{figure}
    \centering
    \includegraphics[width=\columnwidth]{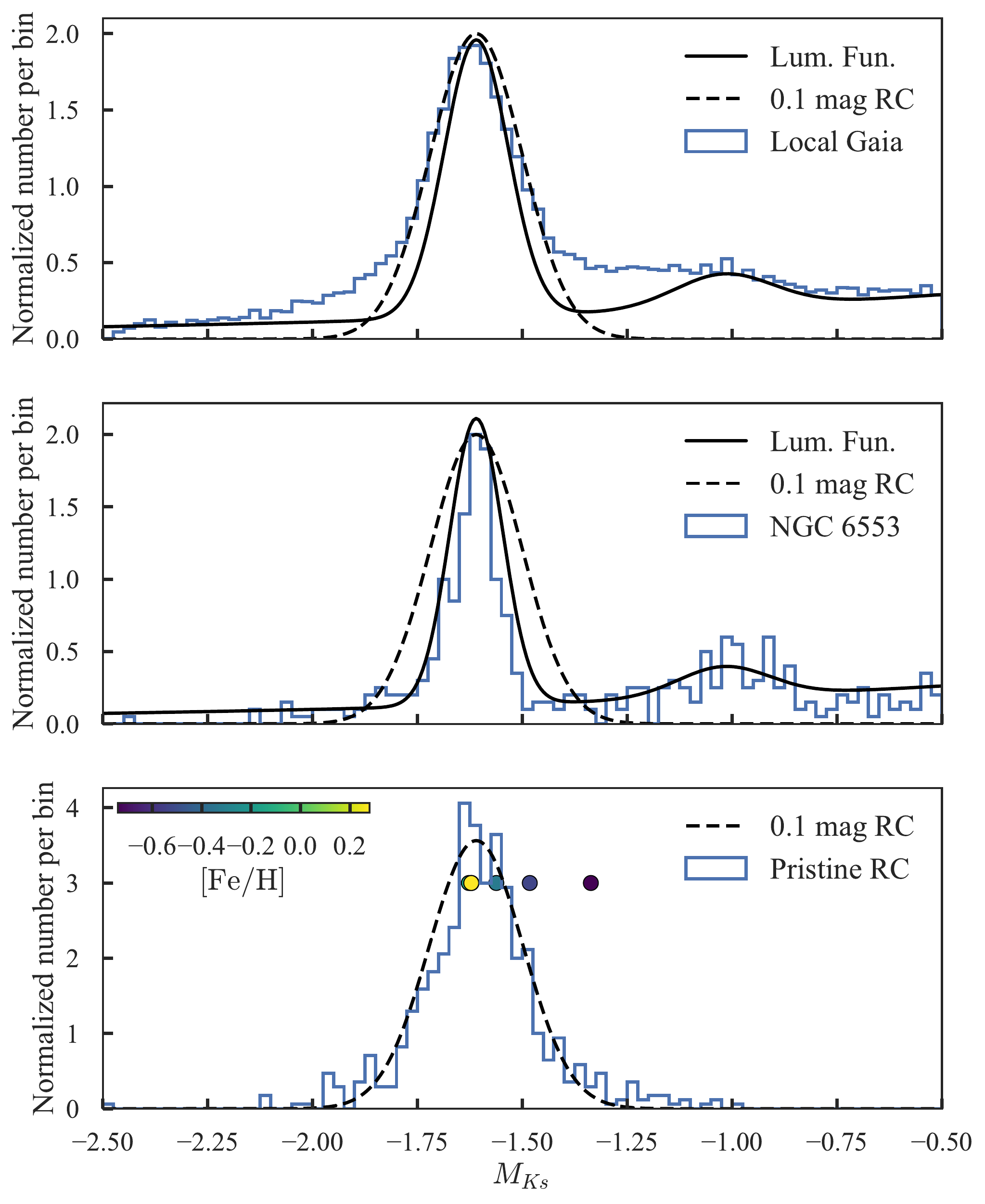}
    \caption{Absolute $K_s$ distributions: the top panel shows stars within $\sim1\,\mathrm{kpc}$ from Gaia DR2, middle panel stars in NGC 6553 and bottom panel just those Gaia DR2 stars identified as `Pristine RC' from \protect\cite{Ting2018}. The black curve is our luminosity function from Paper I and dashed a Gaussian of width $0.1\,\mathrm{mag}$ both convolved by the uncertainty of each dataset. The dots in the bottom panel show the mean magnitude in different metallicity bins.}
    \label{fig:lf}
\end{figure}

We plot the distributions of absolute $K_s$ magnitude for these three datasets in Fig.~\ref{fig:lf} using inverse parallax as a distance estimate. We apply a zero-point parallax offset of $0.05\,\mathrm{mas}$ and assume a distance modulus to NGC 6553 of $13.905$, consistent with the distance of $6\,\mathrm{kpc}$ reported in \cite{Harris1996}. We plot the luminosity function from Paper I and a Gaussian of width $0.1\,\mathrm{mag}$, which have both been convolved by the median uncertainty arising from photometric error, $10\percent$ $\mathrm{E}(B-V)$ error and parallax uncertainty (assuming a systematic floor of $0.021\,\mathrm{mas}$ and a $8\percent$ larger uncertainty than reported, as suggested on the Gaia webpages). There is additional spread from the variation in the parallax offset within the sample. We assume that the data distributions directly give the luminosity function although there are some small incompleteness effects not accounted for. We see that the red clump width for the local stars is broader than our default luminosity function but is well fit by the $0.1\,\mathrm{mag}$ Gaussian. There are broader wings with the fainter wing connecting onto the red giant branch bump. This faint wing is due to secondary red clump stars which are associated with young ($\lesssim1\,\mathrm{Gyr}$) populations \citep{Girardi1999}. The bulge is observed to consist primarily of old stars \citep{Barbuy2018} so this population will not contribute. However, the metallicity distribution of the bulge is broader than the local disc population.

We also compare our luminosity function to that of NGC 6553. Again we convolve by the uncertainty in $K_s$ and $10\percent$ in $\mathrm{E}(B-V)$. We see NGC 6553 has a red clump width similar to our default luminosity function and the red giant branch bump in the correct place. Assuming NGC 6553 is a single age and metallicity population, it seems the width of the red clump in $K_s$ is $\sim0.05\,\mathrm{mag}$. Age and metallicity effects broaden this further. We can assess this somewhat using the `Pristine RC' stars from \cite{Ting2018} which appear to neatly follow the Gaussian with width $0.1\,\mathrm{mag}$. Splitting by metallicity we find the median absolute magnitude increases with decreasing metallicity with a gradient of $-\sim0.5\,\mathrm{mag}/\mathrm{dex}$.
In Paper I we used a vertical gradient in the absolute magnitude of the red clump of $\sim0.1\,\mathrm{mag}/\mathrm{kpc}$ which translates into a metallicity gradient of $\sim-0.03\,\mathrm{dex}/\mathrm{kpc}$ consistent with spectroscopic metallicity gradients observed in the bulge \citep{Barbuy2018}. The fat tail to brighter magnitudes is not seen in the `Pristine RC' sample suggesting these are not red clump stars but background red giant branch.

In conclusion, we have found that the luminosity function from Paper I is similar to that observed in the bulge globular cluster NGC6553 but has too narrow a red clump peak to match the local data from Gaia which points towards a width of $0.1\,\mathrm{mag}$. By modelling the Gaia parallax systematics for an asteroseismic sample, \cite{Hall2019} have argued that the intrinsic width of the red clump in $K_s$ is significantly narrower at $\sim0.03\,\mathrm{mag}$ suggesting either there are additional uncertainties for our local sample bloating the spread or the asteroseismic sample is age and metallicity biased. The local population is not necessarily a reflection of what is expected in the bulge as there are different metallicity distribution widths (the local `Pristine RC' sample has metallicity width $\sim0.24\,\mathrm{dex}$ whilst the bulge has width $\sim0.4\,\mathrm{dex}$, \cite{Hill2011}) and different age distributions (the local distribution has stars of a broader range of ages than observed in the bulge). These two effects will compensate for each other so it is difficult to truly estimate the bulge luminosity function. However, a red clump width between $0.06$ and $0.1\,\mathrm{mag}$ seems appropriate. 

Repeating the pattern speed modelling for $\ell>2.5\,\mathrm{deg}$ using a broader red clump width of $0.12\,\mathrm{mag}$ and the tight priors gives a near identical estimate of the pattern speed of $(41.72\pm2.93)\,\mathrm{km\,s}^{-1}\mathrm{kpc}^{-1}$ (compared to $(42.09\pm2.50)\,\mathrm{km\,s}^{-1}\mathrm{kpc}^{-1}$ of Table~\ref{Table::results}). Therefore, reasonable changes in the luminosity function do not produce significant changes to our analysis.

\bsp	
\label{lastpage}
\end{document}